\documentclass[sigconf]{acmart}
\usepackage{graphicx}
\usepackage{hyperref} 
\usepackage{subfigure}
\usepackage{natbib}
\setcitestyle{numbers, sort}
\usepackage{multirow}
\usepackage{color}
\hypersetup{hidelinks} %hide links of ref
\usepackage{booktabs} % table lines
\usepackage{amsmath,amsthm,amsfonts}
\usepackage[ruled,linesnumbered]{algorithm2e}
\usepackage{balance}

\AtBeginDocument{%
}

\copyrightyear{2024} 
\acmYear{2024} 
\setcopyright{rightsretained} 
\acmConference[CIKM '24]{Proceedings of the 33rd ACM International
Conference on Information and Knowledge Management}{October 21--25,
2024}{Boise, ID, USA}
\acmBooktitle{Proceedings of the 33rd ACM International Conference on
Information and Knowledge Management (CIKM '24), October 21--25, 2024,
Boise, ID, USA}
\acmDOI{10.1145/3627673.3679564}
\acmISBN{979-8-4007-0436-9/24/10}

\author{Jinkun Han}
\affiliation{
  \department{Department of Computer Science}
  \institution{Georgia State University}
  \city{Atlanta}
  \state{Georgia}
  \country{USA}
}
\email{ hjinkun1@student.gsu.edu}

\author{Wei Li}
\affiliation{
  \department{Department of Computer Science}
  \institution{Georgia State University}
  \city{Atlanta}
  \state{Georgia}
  \country{USA}
}
\email{wli28@gsu.edu}

\author{Yingshu Li}
\affiliation{
  \department{Department of Computer Science}
  \institution{Georgia State University}
  \city{Atlanta}
  \state{Georgia}
  \country{USA}
}
\email{yili@gsu.edu}

\author{Zhipeng Cai}
\affiliation{
  \department{Department of Computer Science}
  \institution{Georgia State University}
  \city{Atlanta}
  \state{Georgia}
  \country{USA}
}
\email{zcai@gsu.edu}

\renewcommand{\shortauthors}{Jinkun Han, Wei Li, Yingshu Li, \& Zhipeng Cai}

\begin{document}
\begin{sloppypar}

\title{Quantum Cognition-Inspired EEG-based Recommendation via Graph Neural Networks}

% ================================ author ==================================

%================================ abstract ==================================
\begin{abstract} 
Current recommendation systems recommend goods by considering users' historical behaviors, social relations, ratings, and other multi-modals. 
Although outdated user information presents the trends of a user's interests, no recommendation system can know the users' real-time thoughts indeed.
With the development of brain-computer interfaces, it is time to explore next-generation recommenders that show users' real-time thoughts without delay.
Electroencephalography (EEG) is a promising method of collecting brain signals because of its convenience and mobility.
Currently, there is only few research on EEG-based recommendations due to the complexity of learning human brain activity.
To explore the utility of EEG-based recommendation, we propose a novel neural network model, QUARK, combining Quantum Cognition Theory and Graph Convolutional Networks for accurate item recommendations.
Compared with the state-of-the-art recommendation models, the superiority of QUARK is confirmed via extensive experiments.
\end{abstract}

%================================ ccs ==================================
\begin{CCSXML}
<ccs2012>
   <concept>
       <concept_id>10002951.10003317.10003347.10003350</concept_id>
       <concept_desc>Information systems~Recommender systems</concept_desc>
       <concept_significance>500</concept_significance>
    </concept>
   <concept>
       <concept_id>10002951.10003317.10003347.10003352</concept_id>
       <concept_desc>Information systems~Information extraction</concept_desc>
       <concept_significance>500</concept_significance>
    </concept>
 </ccs2012>
\end{CCSXML}

\ccsdesc[500]{Information systems~Recommender systems}
\ccsdesc[500]{Information systems~Information extraction}

%================================ keywords ==================================
\keywords{Recommendation System, Quantum Cognition Theory, Graph Neural Network, EEG Data, Personalization}

\maketitle
%============================= Introduction =====================================

\section{Introduction}
\begin{figure}
   \centering
   \includegraphics[width=0.5\textwidth]{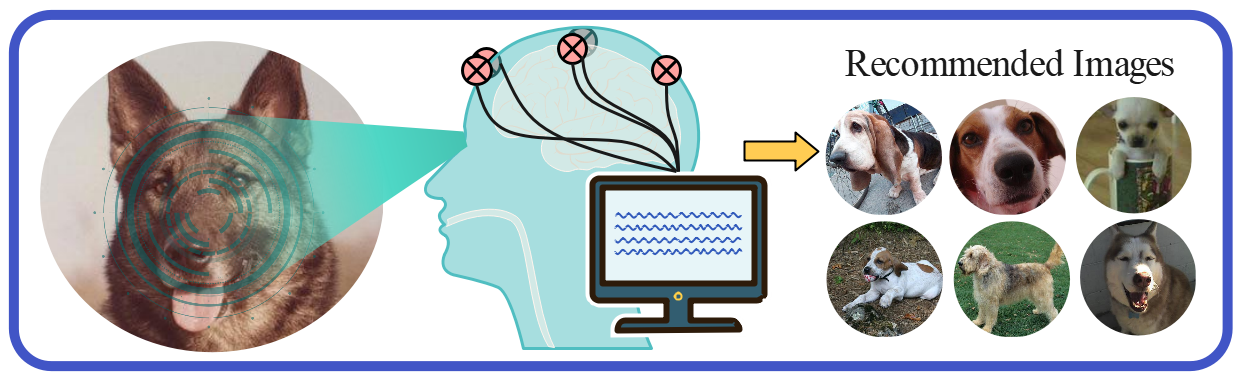}
   \caption{Understand the thoughts or needs, and recommend related items.}
   \label{fig:introduction}
\end{figure}

Brain-computer interface (BCI) is an emerging field attracting research institutions and industries in the areas of motor imagery classification~\cite{mwata2022improving}, emotion recognition~\cite{li2022eeg}, disease diagnosis and detection~\cite{maitin2022survey}, music imagery~\cite{mahmood2022effect, qian2022deep}, and other tasks~\cite{apicella2022eeg}.
As one of the BCIs, non-invasive electroencephalography (EEG) has become popular and is commonly used own to its convenience and mobility.
EEG measures electrical activities in human brains and reflects people's real-time thoughts and feelings about something. 
Although thought is a very vague concept, EEG digitizes thoughts by capturing the information indicating individuals' reflections about what they see, think, feel, hear, {\em etc.} 
Such learnt ``thoughts" about items then can be treated as inputs to predict what to recommend as shown in Figure~\ref{fig:introduction}. 
%Usually, EEG data is continuously collected when users are experiencing the EEG-embedded services ({\em e.g.} VR games).
%While, the users may not want to be recommended all the time, so detecting when to recommend items important, which, however, is another research problem. 
In this paper, we focus on the utilization of EEG signals for item recommendation, which has not been addressed in literature.

Over the past decades, recommendation systems have become society's critical infrastructures in retail, online shopping, entertainment, advertising, finance, {\em etc.} against the era of information explosion~\cite{survey1,survey2}, in which historical information (such as user behaviors, attributes of the items, social relationships, and multimodals~\cite{example1,example3}) is commonly exploited to predict user preference. 
Although historical information depicts the trends of a user's behavior, a user may change the mind from time to time. 
Therefore, it is in the desired need of a real-time thinking-based recommendation method for the next-generation recommendation -- EEG-based recommendation will be able to fill the research blank as well as bring a brand new enhanced user experience to advance personalized service provision. For examples, 
(i) personalized product/Ad placement based on user thoughts in VR games~\cite{9660742, seminati2022multisensory} when a user is looking at some products; 
(ii) friends or instructions recommendation in augmented reality~\cite{lee2018metasurface} when a person has thoughts on physical items; 
(iii) online shopping to skip search~\cite{guo2013new} as the experience of ``freshness" is always the primary consumption force; 
and 
(iv) music/movie recommendation~\cite{guo2012eeg} using the truest feelings. 

Moreover, different from traditional data, EEG signals contain mixed brain reflections that are hard to be split from the signals, where past thoughts unavoidably influence future thoughts.
It is essentially important and difficult to learn what composes the mixture of brain reflections and how past thoughts influence future thoughts.
Unfortunately, the state-of-the-art works process EEG data the same as traditional data~\cite{khademi2022transfer,maitin2022survey,ahmad2022eeg} without considering the features of mixed brain reflections.
To deal with these challenges, we employ Quantum Cognition Theory (QCT)~\cite{busemeyer2012quantum} to pre-process EEG signal before inputting it into the recommendation scheme. 
In particular, QCT can decompose ``thoughts" into latent factors if a quantum space is given, and quantum interference can find the relationship between past thoughts and future thoughts.

As a pioneer in exploring the feasibility of ``recommend what you think", in this paper we propose a model, {\bf QU}antum Cognition-Inspired EEG-based Recommendation by Using gr{\bf A}ph Neu{\bf R}al Networ{\bf K} (QUARK), which integrate Quantum Cognition Theory and Graph Convolutional Networks (GCN) to recommend items based on EEG signals. 
Our QUARK model has multi-fold technical innovations: 
(i) EEG data collected within a period contains changes of thoughts, so a sliding window method is designed to segment EEG data to characterize the sequential property of thoughts in the temporal domain; 
(ii) as each thought is a mixture of thinking, QCT is utilized to decompose the thought via applying quantum theory to model cognitive phenomena; and 
(iii) GCN aggregates the generated EEG information to a final representation for recommendations by using continuity and interference graphs that demonstrate how past thoughts influence future thoughts.
In summary, the major contributions of this paper are addressed below:
1) To the best of our knowledge, this is the first framework that discloses a detailed EEG-based item recommendation with neural network, Quantum Cognition Theory, and Graph Convolutional Networks.
2) QUARK is proposed to recommend items, taking into account the change of thoughts, decision-making from thoughts, continuity and interference between thoughts. 
3) Extensive experiments are well set up on real data, and the results validate that QUARK outperforms the classic recommendation models not only in top $k$ recommendation but also in feeling and style detection.

The rest of this paper is organized as follows. 
Section~\ref{RelatedWork} introduces the related work.
Section~\ref{Preliminaries} introduces some preliminaries.
QUARK is proposed in Section~\ref{Methodology}. 
After evaluating QUARK in Section~\ref{Evaluation}, we conclude this paper in Section~\ref{Conclusion}.
In addition, We provide supplemental materials~\footnote{ \url{https://github.com/KK429312/CIKM2024Appendix/blob/main/QuantumEEG_240805_JKH_CIKM_Appendix.pdf}\label{outside_link}}.

\begin{figure*}
    \centering
    \includegraphics[width=0.95\textwidth]{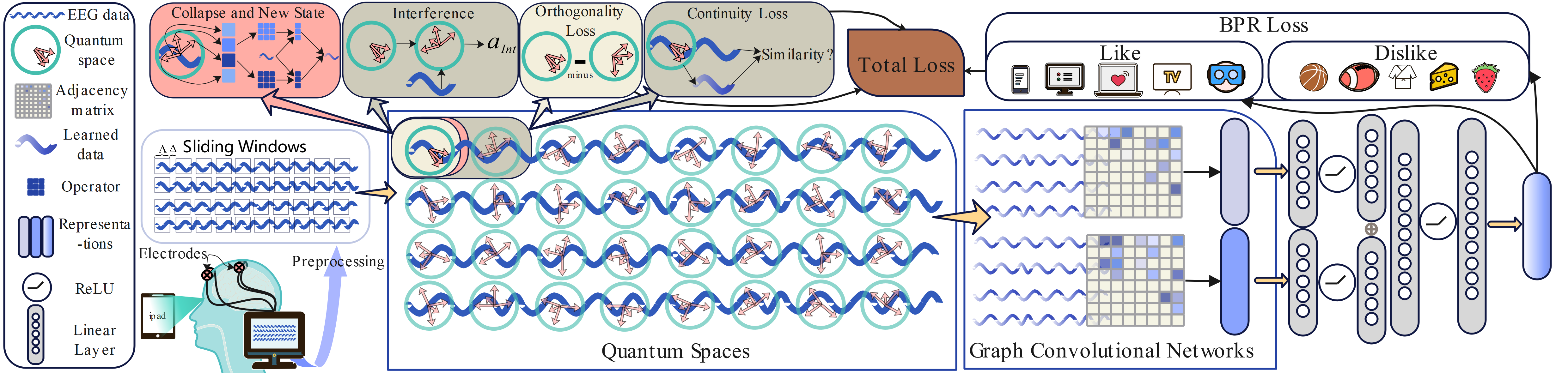}
    \caption{The framework of QUARK.}
    \label{fig:framework}
\end{figure*}
%

%==================================================================
\section{Related Work} \label{RelatedWork}
% EEG signals are collected from brain-computer interference (BCI) with electrodes~\cite{fouad2015brain}.
% EEG is convenient, highly secure, and non-invasive compared to other BCI methods, including fMRI, MEG, NIRS, PET, and EROS~\cite{7079068}.
% Therefore, EEG is likely to become the most popular wearable technology for collecting brain signals in the future.
%In brain signal processing, understanding the content of the brain signal is essentially important.

Traditionally, items are recommended using matrix factorization methods, where user-item ratings are computed according to the user and item matrices~\cite{koren2009matrix}. 
Rendle {\em et al.} developed a personalized recommendation optimization method based on Bayesian theory in 2009~\cite{bpr}.
With the development of neural networks and deep learning, Guo {\em et al.} proposed DeepFM that takes advantage of hidden layers to extract latent features for the recommendation~\cite{ijcai2017239}. 
He {\em et al.} used neural networks and designed a neural collaborative filtering framework to learn user-item interaction in 2017~\cite{he2017neural}.
Later, the personalized recommendation is considered.
Naumov {\em et al.} from Facebook~\cite{naumov2019deep} proposed DLRM to explore personalized recommendation and allow data parallelism to scale-out computation.

The brain-computer interface has been used to collect EEG signals in many emerging applications. 
To understand the contents of brain signals when processing them, quantum neurobiology has been proposed by Swan {\em et al.}~\cite{swan2022quantum} to study brain contents via quantum theory.
Busemeyer {\em et al.} reveal that quantum cognition theory can help understand the sequential context of how previous thoughts influence the next~\cite{busemeyer2015quantum}.
Bruza {\em et al.} proved quantum cognition-based human reasoning is more effective than traditional probabilistic models. 
Moreover, quantum theory combined with deep learning brings a new perspective on learning representation.
%In 2005, Karayiannis {\em et al.} proposed a quantum neural network (QNN) with EEG to detect epileptic seizures~\cite{karayiannis2006evaluation}. 
%Aljazaery {\em et al.} designed a QNN with EEG to classify five classes in 2011~\cite{aljazaery2011classification}.
In~\cite{taha2018eeg}, Taha {\em et al.} proposed QRNN-AR that possesses high classification capabilities thanks to quantum recurrent neural networks.
Quantum machine learning designed by Li {\em et al.} is used to extract EEG features and perform the classification~\cite{9112355}.

Currently, no EEG-based framework has taken the advantage of integrating QCT and EEG on marketing, although neuromarketing is crucial for the advertising and promotion market of approximate 400 billion dollars~\cite{9377473}.
%Also, computer scientists cover all the fields of daily life recommendation~\cite{ko2022survey} but have not yet paid attention to EEG-based recommendation.
Therefore, to fill this blank, we propose a novel model, QUARK, to capture people's instant thoughts through EEG data based on QCT for item recommendation.

%The most recent advanced recommendation technologies consider graph neural networks~\cite{ojo2022visgnn,zhang2022dynamic}, because the multiple relations in graphs guarantee the efficiency and precision of personalized recommendation models.
%Since EEG-based recommendation is a brand new area that we do not reach the stage of considering the social or user relations in graphs, so we do not introduce more works about the graph-based recommendation.
%
%
\begin{table}
\caption{Notations (\# denotes the number).}
\label{table:variables}
\begin{center}
\centering
\scalebox{0.7}{
\begin{tabular}{cccc}
\toprule
Variables & Description & Variables & Description\\
\midrule
    $X$ & Set of EEG data & $Y$ &  Set of item data\\
    $M$ & \# of electrodes & $N$ & Length of any EEG signal\\
    $\mho$ & \# of segments of a signal &$\Phi$ & Set of EEG segments\\ 
    $\Lambda$ & Width of sliding windows & $\Delta$  & Length of sliding steps\\
    $\Theta$ & Set of events & $B$ & Set of basis vectors\\
    $O$& Set of Quantum operators &  $c$ & \# of selected basis vectors\\
    $\epsilon$  & Index set of selected basis vector & $P$ & Set of probabilities\\
    $ \widehat{A}$ &  Continuity matrix & $\widetilde{A}$ & Interference matrix\\
    $H$ & Linear layer & $S$ & Mask matrix\\
    $D$ &  Total depth of GCN & $\Gamma $ & List of learnable parameters\\
    $K$ & \# of sampled items & $k$ & \# of recommended items\\ 
\bottomrule
\end{tabular}
}
\end{center}
\end{table}

\section{Preliminaries}\label{Preliminaries}
%In this paper, the frequently used notations are described in Table~\ref{table:variables}. The necessary preliminaries are briefly summarized below.

%\subsection{Hilbert Space and State} \label{Hilbert_Space_and_State}
Hilbert space is a vector space with inner product operations and is composed of a set of abstract points. 
% QCT lies in Hilbert space with Property 1.

\textbf{Property 1.} \textit{ The dot product in Hilbert space satisfies: $(o_1 + o_2)    o_3 = o_1  o_3 + o_2   o_3$ for any vector $o_1$, $o_2$, and $o_3$.}

Following Busemeyer's~\cite{busemeyer2012quantum} definition, each point in Hilbert space is called a vector denoted as $|x \rangle$ or $|b \rangle$ by a ket notation $|\cdot \rangle$, which are the state and basis vector in this paper, respectively. A bra notation is defined as $\langle \cdot |$ representing the transpose of ket.
QCT postulates a unit-length state vector $|x \rangle$ as the state vector in Hilbert space, where $\langle x|x \rangle=1$ with inner product $\langle \cdot|\cdot \rangle$. 
This state vector describes a person's thought in limited time steps.  

\textbf{Definition 1.} \textit{As an EEG vector has finite dimensions and no complex numbers, an EEG vector in Hilbert space with the bra-ket notation can be transformed into the Euclidean space~\cite{darmochwal1991euclidean} with the same value only by removing the bra-ket notation. 
In contrast, unit length must be guaranteed from Euclidean space to Hilbert space.
}

%\subsection{Events and Measurement} \label{Events_and_Measurement}
A segment of EEG data with fixed size is named an event. Each event in QCT has a finite dimension $|B_{m,i}|$, where $B_{m,i}\subseteq B$ is a set of basis vectors with index $m$ and $i$.
Each element of $B_{m,i}$ represents a kind of choice.
Hence, a set of basis vectors (a set of choices), where $B_{m,i} = \{|b_j\rangle,j = 1,2,\dots,|B_{m,i}|\}$, spans the quantum space (the event).  
Furthermore, QCT provides a way to measure decision making if a person is in a state $|x\rangle$, which is $p=\left \| \langle b_{j}|x\rangle  \right \|^2$ named collapse, where $p$ is the probability (a scalar) calculated from a basis vector and $b_j\in B_{m,i}$.
The result $p$ reveals how possible it is to make a decision on choice $b_j$, while an extended measurement can be utilized if a new state is required, that is, an operator $o_i$, which projects the old state to a new state. 
To obtain a new state, QCT applies $|x_{new}\rangle=o_j|x\rangle$, where $o_j=|b_j \rangle\langle b_j|$ is the operator of basis vector $b_{j}$ with the outer product $|\cdot \rangle\langle \cdot|$.
To deeply mine a person's thought, pure state and mixed state are defined in Definition 2.

\textbf{Definition 2.} \textit{The new state collapsing along one basis vector ($|x_{new}\rangle=o_i|x\rangle$) is called pure state, while collapsing along two or more than two basis vectors is named mixed state, formally named a superposition state in QCT, with $|x_{new}\rangle=\sum{o_j|x\rangle}$ and selected projectors $\{ o_j | selected \; j\}$. 
}

%\subsection{Quantum Interference}\label{Quantum_Interference}
If the past event $\Theta_{m,1}$ influences the future event $\Theta_{m,2}$, then QCT does not obey the distribution law used in classic probability theory (CPT)~\cite{jaynes2003probability} ({\em i.e.} $p(\Theta_{m,2})=p(\Theta_{m,2})(p(\Theta_{m,1})+p(\bar{\Theta}_{m,1}))=p(\Theta_{m,2})p(\Theta_{m,1})+p(\Theta_{m,2})p(\bar{\Theta}_{m,1})$ with $\bar{\Theta}_{m,1}$ representing the event $\Theta_{m,1}$ not occurring), because CPT works only if two events are independent, while QCT has the superiority of mathematically measuring the interference between two events that are not independent.
Theorem.~\ref{theorem:interference} defines the interference value $\eta_{\Theta_{m,1}  \rightarrow \Theta_{m,2}}$ when event $\Theta_{m,1}$ interferes with event $\Theta_{m,2}$ with the state $|x\rangle $ in $\Theta_{m,2}$ (See Supplement [\ref{outside_link}] Section 1 for proof). In this paper, each event can be regarded as a segment of EEG data that semantically equals a mixed thought.
\begin{theorem}
\label{theorem:interference}
If past event $\Theta_{m,1}$ influences the future event $\Theta_{m,2}$, the operator interpreting the event $\Theta_{m,1}$ occurring is defined as $o_{\Theta_{m,1}}=\sum {|b_j \rangle\langle b_j|}$ with selected basis vectors $\{b_j| selected \; j\}$, while the operator of $\Theta_{m,1}$ not occurring is defined as $o_{\bar{\Theta}_{m,1}}=\sum {|b_{\bar{j}} \rangle\langle b_{\bar{j}}|}$ with selected basis vectors $\{b_{\bar{j}}| selected \; \bar{j}\}$.
The operator interpreting the event $\Theta_{m,2}$ occurring is defined as $o_{\Theta_{m,2}}=\sum {|b_\kappa \rangle\langle b_\kappa|}$ with selected basis vectors $\{b_\kappa| selected \; \kappa\}$, where $j$, $\bar{j}$, and $\kappa$ are the indexes of basis vectors. Then, the interference value is defined as: %by Eq.~\ref{eq:Preliminaries_interference}.
%\begin{equation}
%\label{eq:Preliminaries_interference}
$\eta _{\Theta_{m,1}  \rightarrow  \Theta_{m,2}} = 2 \times \langle x| o_{\bar{\Theta}_{m,1}} o_{\Theta_{m,2}} o_{\Theta_{m,2}} o_{\Theta_{m,1}}|x\rangle$ 
%\end{equation}
\end{theorem}

%==================================================================
\section{Methodology} \label{Methodology}
In this paper, the frequently used notations are described in Table~\ref{table:variables}. 
In this section, we proposed a novel model, QUARK, for item recommendation by utilizing EEG signals, GCN, and QCT.
As shown in Figure~\ref{fig:framework}, QUARK contains 4 major components, including (i) Data Pre-processing \& Sliding Windows, (ii) Quantum Spaces \& New States, (iii) Matrices \& Graph Convolutional Networks, and (iv) Model Training \& Prediction.

The main idea of QUARK is briefly introduced below.
Given any EEG instance $x\in X$ with $M$ electrodes on an EEG device and $N$ time steps, the pre-processed instance $x'$ is produced correspondingly. 
To build graphs capturing the relationships between thoughts, $x'$ is processed based on rows and sliding
window.
QUARK divides $x'$ according to the width of the sliding window $\Lambda\in \{ 1,2,\cdots,N \}$, indicating how much data is included in a single segment, and the sliding size $\Delta \in \{1,2,\cdots,N\}$, representing how much the window moves forward on the EEG signal, to obtain a set of segments $\Phi$.
%where $m$ is the index of $m$-th row of $x'$ generated from $m$-th electrode.
Then, the segmented EEG data $x_{m,i}^* \in \Phi$ is the $i$-th segment of $m$-th row of $x'$ ($1\leq m \leq M$) and fed into a quantum space to learn a new EEG signal $x_{m,i}^{\circ}$ that is a mixed state in quantum space. 
Meanwhile, the continuity adjacency $\widehat{A}$ and the interference adjacency $\widetilde{A}$ are generated by considering the continuity of EEG signals and quantum interference.
Both $\widehat{A}$ and $\widetilde{A}$ are used by GCN to aggregate $x_{m,i}^{\circ}$ to form two aggregated representations $\widehat{x}^{\blacksquare}$ and $\widetilde{x}^{\blacksquare}$, respectively.
Finally, $\widehat{x}^{\blacksquare}$ and $\widetilde{x}^{\blacksquare}$ are concatenated and fed into linear layers to obtain the final representation $\overline{x}$.
QUARK samples $K$ items $\{y_1, y_2, \dots, y_K \}$ to compute similarity scores $\{\overline{x}y_1,\overline{x}y_2,\cdots,\overline{x}y_K  \}$ and then recommends top-$k$ items with $k$-highest scores.
\begin{algorithm}
\caption{Sliding Window}
\label{alg:Sliding_Window}
\KwIn{data $x'$, windows size $\Lambda$, sliding size $\Delta$}
\KwOut{ the set of segments $\Phi$}
  $\Phi$ = $\{\}$, $m=1$, $i=1$\;
  \For{$m \leq M$}
  {
    \For{$i \leq \mho$}
    {
        $x_{m,i}^* = x'[m][(i-1) \times \Delta + 1: (i-1) \times \Delta  + \Lambda + 1 ]$\;
        $\Phi.add(x_{m,i}^* )$\;
    }
  }
\KwResult{$\Phi$}
\end{algorithm}

%---------------------------------------------------------------------------------
\subsection{Data Pre-processing \& Sliding Windows}
%\subsubsection{Data Pre-processing}

Data scaling technique is used to speed up the training process and avoid gradient explosion in deep learning~\cite{NEURIPS2018_905056c1}, especially for EEG data that does not have the same range of values. 
In Eq.~\eqref{eq:mean_norm}, mean normalization is considered because it scales the EEG data to a suitable range $(-1, 1)$ and keeps the raw data pattern. 
\begin{equation}
\label{eq:mean_norm}
x'=\frac{x-mean(x)}{max(x)-min(x)}.
\end{equation}
%where $x'$ is the preprocessed EEG data, $x$ is the raw EEG data with $x \in \mathbb{R}^{M\times N}$ denoting the $M\times N$ matrix, and $M,N$ are the number of electrodes and time steps, respectively.

%\subsubsection{Sliding Windows}
Typically, EEG data contains thinking factors, such as imagination, association, decision-making, inference, judgment, reasoning, and feeling, {\em etc}.
They happen sometimes simultaneously and sometimes sequentially and are significant in identifying a person's thoughts.
As a state that contains the factor(s) generally lasts for a short time, a sliding window-based method is designed in QUARK to segment human states in Algorithm~\ref{alg:Sliding_Window}, where the total number of segments is $|\Phi|=M\times \left \lfloor \frac{N-\Lambda+\Delta}{\Delta} \right \rfloor$, the number of segments of the $m$-th electrode is $\mho = \left \lfloor \frac{N-\Lambda+\Delta}{\Delta} \right \rfloor$. 
In line 4 of Algorithm~\ref{alg:Sliding_Window}, the $m$-th row of $x'$ and the elements between columns $(i-1) \times \Delta + 1$ and $(i-1) \times \Delta  + \Lambda + 1$ are selected to generate $x_{m,i}^*$ representing a super-short-term thinking that contains multiple thinking factors.
%All elements $x_{m,i}^*\in \mathbb{R}^{\Lambda\times 1}$ in $\Phi$ make up the segmented matrix $x^*\in \mathbb{R}^{M\times \mho \times \Lambda}$.
These factors are helpful to learn the user's current needs and trends and can be exploited to make an accurate recommendation.

%

%-------------------------------------------------------------------------
\subsection{Quantum Spaces \& New States}
To learn the EEG representation from mixed states and uncertain states, quantum spaces are utilized to simulate the human decision-making process.
QCT is an effective method of decomposing states onto basis vectors to find the decisions of the thoughts in quantum space.
A basis vector represents a latent thinking factor that might be one of imagination, association, decision-making, inference, or other unknown factors influencing human thoughts. The number of thinking factors can be more than four, so this paper does not use traditional $|00\rangle$,$|01\rangle$,$|10\rangle$, and $|11\rangle$ as basis vectors. Instead, learnable vectors $b_{j}$ are used. 
The meaning of thinking factors does not affect the learning process and recommendation results, so there is no need to know their exact meanings in our model.
Also, explaining the thinking factors is out of the scope of this paper, which could be further studied in our future work.
From the previous step, 
$|\Phi|$ quantum spaces are generated, each of which is regarded as one latent event $\Theta_{m,i}$ containing $|B_{m,i}|$ latent thinking factors as basis vectors, where $B_{m,i}\in \mathbb{R}^{|B_{m,i}|\times \Lambda}$ is a randomly generated matrix and $B=\{ b_j|b_j\in B_{m,i},1\leq m\leq M,1\leq i\leq \mho \}$. 

For every segmented $x_{m,i}^*$, the unit length revise function in Eq.~\eqref{eq:revise1} is applied to obtain a quantum space vector $|x_{m,i}^* \rangle$ 
and Eq.~\eqref{eq:qm_prob} is used to form the set of the probabilities of latent thinking factors, where $\left \| b_j| x_{m,i}^*\rangle  \right \|^2$ is the probability that $|x_{m,i}^* \rangle$ collapses along basis vector $b_j$.
This probability represents how possible the state ({\em i.e.}, thought) $|x_{m,i}^* \rangle$ is composed by the thinking factor $b_j$.
Then, the indices of the $c$-highest probabilities in $P_{m,i}$ are picked up to form an index set $\epsilon_{m,i}^c$ for further computing.
\begin{equation}
\label{eq:revise1}
|x_{m,i}^* \rangle = \frac{x_{m,i}^*}{\sqrt{\sum_{ j=1}^{\Lambda}(x_{m,i,j}^*)^2 }},
\end{equation}
\begin{equation}
\label{eq:qm_prob}
P_{m,i} =  \{ \left \| b_j| x_{m,i}^*\rangle  \right \|^2 | b_j \in B_{m,i} \},
\end{equation}
where $x_{m,i,j}^*$ is the $j$-th element of vector $x^*_{m,i}$ .
Due to the selection of probabilities, QUARK can be treated as a variant of ensemble learning~\cite{sagi2018ensemble,dong2020survey} and dropout~\cite{baldi2013understanding}. Ensemble learning uses some different models to predict the data, gets corresponding predictions, and then votes based on all predictions to make a final prediction, which usually outperforms the learning of a single model but takes time to design the models manually. 
Dropout is developed to avoid the manually designed models by randomly dropping some neurons in neural networks, so the prediction may be dependent on random seeds.
QUARK combines the advantages of ensemble learning and dropout naturally, since it provides any EEG data instance with $\begin{pmatrix} |B_{m,i}|\\c \end{pmatrix}^{|\Phi|}$ path options utilized as models just as in ensemble learning, is self-adaptive to select the most suitable one without voting, and does not randomly drop the paths. 
The index set $\epsilon_{m,i}^c$ reveals that a person's state $|x_{m,i}^*\rangle$ collapses towards a few basis vectors; that is, the current thought makes some decisions to form the next thought. 
Note that a person is hardly in a pure state due to the various emotions and environments, any newly generated state,  $|x_{m,i}^{\circ}\rangle$, is usually a mixed state and can be expressed by Eq.~\eqref{eq:new_state}.
\begin{equation}
\label{eq:new_state}
|x_{m,i}^{\circ}\rangle =\sum_{j\in \epsilon_{m,i}^c}{o_{j}|x_{m,i}^{*}\rangle} =\sum_{j\in \epsilon_{m,i}^c}{|b_j\rangle \langle b_j|x_{m,i}^{*}\rangle},
\end{equation}
where $o_j$ is the operator computed by outer product $|b_j\rangle \langle b_j|$.
Because of  Property 1, Eq.~\eqref{eq:new_state} can be accelerated by reducing the matrix multiplication through Eq.~\eqref{eq:new_state2}, especially when the number of selected basis vector, $c$, is large. The matrix $x^{\circ}\in \mathbb{R}^{M\times \mho \times \Lambda}$ is obtained to represent the learned EEG data.
\begin{equation}
\label{eq:new_state2}
|x_{m,i}^{\circ}\rangle=(o_1+o_2+\dots+o_{|\epsilon_{m,i}^c|})|x_{m,i}^{*}\rangle=(\sum_{j\in \epsilon_{m,i}^c}{|b_j\rangle \langle b_j|})|x_{m,i}^{*}\rangle.
\end{equation}
% To be more clear, $|x_{m,i}^{*}\rangle$ collapses towards selected latent thinking factors in this latent event $\Theta_{m,i}$, such as association, reasoning, and judgment, where each $|b_j\rangle \langle b_j|x_{m,i}^{*}\rangle$ represents how the $|x_{m,i}^{*}\rangle$ decomposed along basis vector $b_j$ influences the generation of following thought $|x_{m,i+1}^{*}\rangle$.
% Hence, this kind of influence is the basis that QUARK is optimized by continuity loss in Section~\ref{model_training_and_prediction}.

%, where $M,\mho$ are the first dimension and second dimension, and $\Lambda$ is the length of $x_{m,i}^{\circ}$.

%-------------------------------------------------------------------------------------------------------------------------
\subsection{Matrices \& Graph Convolutional Networks}\label{RAMIAM}
In this part, the continuity adjacency matrix $\widehat{A}$ is computed to form new continuous thoughts, and the interference adjacency matrix $\widetilde{A}$ is constructed to simulate the influence process from previous thoughts to future thoughts. 
With $\widehat{A}$ and $\widetilde{A}$, a GCN aggregator~\cite{KlicperaBG19,wu2020connecting} is utilized to learn graph representations. 
%Meanwhile, two masks are applied to avoid the influence of future thoughts on past thoughts. 

\subsubsection{Continuity Adjacency Matrix }
The dot similarity in Euclidean space is computed as the weights of continuity adjacency matrix $\widehat{A}\in \mathbb{R}^{|\Phi|\times |\Phi|}$ in Eq.~\eqref{eq:relation_matrix}.
\begin{equation}
\label{eq:relation_matrix}
\widehat{A} =  \Psi(x^{\circ},(|\Phi |,\Lambda )) \Psi (x^{\circ},(\Lambda, |\Phi|)), 
\end{equation}
where $\Psi(\cdot,\cdot)$ reshapes $x^{\circ}$ into a new shape $(\cdot,\cdot )$~\cite{pytorch_view}. 
In $\widehat{A}$, a positive weight means two thoughts are similar, a negative weight implies that two thoughts are dissimilar, and a zero weight means there is no similar or dissimilar relation between two thoughts. 
Similar thoughts can help extract a user's preference for recommendations, dissimilar thoughts disturb the capability of learning qualified preference representation,  and non-related thoughts do not influence the preference extraction in GCN method.
Thus, ReLU activation function~\cite{li2017convergence} is applied to drop negative weights, $\widehat{A}' = ReLU(\widehat{A})$.
To deeply investigate the influence of $\widehat{A}'$, a filter ratio $\alpha \in[0,1]$ is introduced to indicate how many weights are kept.
When $\alpha=0$, all values in $\widehat{A'}$ are maintained; when $\alpha>0$, the values smaller than a threshold are filtered out, which implies that the influence of the filtered values on preference extraction are not significant enough for consideration.
Moreover, with a larger $\alpha$, more values can be filtered out, indicating a higher filter ratio.
%{\color{red}
%A self-adaptive range method $\alpha (max (\widehat{A}')-min (\widehat{A}'))+min (\widehat{A}')$ is developed, where different numbers of weights are dropped for different pattern of EEG.
%Selecting the fixed number of weights from low to high is difficult and time consuming, while a self-adaptive selection is easy to implement and neural network can learn to fit as the self-adaptive method can be regarded as a non-linearity function in deep learning.}
Then, the final continuity adjacency matrix $\widehat{A}^{\circ}$ is obtained  in Eq.~\eqref{eq:relation_retention_percentage}, where $\widehat{a}^{\circ}_{j,k} \in \widehat{A}^{\circ}$ ($j=(m-1)\mho +i$ and $k=(v-1)\mho +w$) represents the similarity between the past event $\Theta_{v,w}$ and future event $\Theta_{m,i}$.
\begin{equation}
\label{eq:relation_retention_percentage}
  \widehat{a}^{\circ}_{j,k} = 
  \begin{cases}
    \widehat{a}'_{j,k}, &\text{if $i>w$ and }\\
    &\text{$\widehat{a}'_{j,k} \geq   \alpha (max (\widehat{A}')-min (\widehat{A}'))+min (\widehat{A}'),$}\\
	0, &\text{else,}
  \end{cases}
\end{equation}
where $\widehat{a}'_{j,k} \in \widehat{A}'$, and $i>w$ is required to avoid the influence of future events on past events since a user thinks over time.
The continuity adjacency matrix reveals a person's thoughts in two aspects: (i) it is a similarity matrix to simulate the regrouped thoughts; and (ii) higher weights are calculated when a person repeats the same thought as the past, {\em i.e.}, the repeatability of thoughts.
Particularly, the continuity adjacency matrix only measures continuity and repeatability between $x_{m,i}^{\circ}$ and $x_{m,i+1}^{\circ}$ without the calculation between $x_{m,i}^{\circ}$ and $x_{m,i+1}^{*}$ stated in continuity loss in Section~\ref{model_training_and_prediction}.
The continuity loss means that the learned EEG data $x_{m,i}^{\circ}$ is limited to be similar to pre-processed EEG $x_{m,i}^{*}$, while continuity adjacency matrix shows that all learned EEG data also keep certain continuity.

\subsubsection{Interference Adjacency Matrix}
Formally, $\Theta_{m,i}$ means event $\Theta_{m,i}$ occurs, while $\bar{\Theta}_{m,i}$ means event $\Theta_{m,i}$ does not occur. 
To reduce computation complexity in QUARK, $\bar{\Theta}_{m,i}$ is not computed as the complement of $\Theta_{m,i}$; instead, the set of $c$-lowest index $\epsilon_{m,i}^ {\bar{c}}$ is found from $P_{m,i}$ in Eq.~\eqref{eq:qm_prob} to approximate $\bar{\Theta}_{m,i}$, where $2c\leq |B_{m,i}|$ to ensure $\Theta_{m,i}$ and $\bar{\Theta}_{m,i}$ are disjoint. 
Each interference weight $\widetilde{a}_{j,k}\in \widetilde{A}$ ($j=(m-1)\mho +i$ and $k=(v-1)\mho +w$) is computed by Eq.~\eqref{eq:interference_matrix} based on Theorem~\ref{theorem:interference} with $\widetilde{A}\in  \mathbb{R}^{|\Phi |\times| \Phi |}$. 
\begin{equation}
\label{eq:interference_matrix}
\widetilde{a}_{j,k}  = 2 \times \langle x_{m,i}^{*}| o_{\bar{\Theta}_{v,w}} o_{\Theta_{m,i}}o_{\Theta_{m,i}} o_{\Theta_{v,w}}|x_{m,i}^{*}\rangle.
\end{equation}
Specifically, $\widetilde{a}_{j,k}=0$ if two events are independent, $\widetilde{a}_{j,k}<0$ if the previous event prevents future events from happening, and $\widetilde{a}_{j,k}>0$ if the previous event promotes future events.
Therefore, in $\widetilde{A}$, only the positive values help describe a user's preference.
%
%If the past event $\Theta_{v,w}$ prevents the future event $\Theta_{m,i}$ from happening but $\Theta_{m,i}$ already exists, it is a contradiction to say that  represent the user's preference.
%On the contrary, if the past event $\Theta_{v,w}$ promotes the future event $\Theta_{m,i}$ so that event $\Theta_{m,i}$ exists, the user's preference can be described by them.
%
Accordingly, to remove  the negative in $\widetilde{A}$, the ReLU function is applied to get $\widetilde{A}'=ReLU(\widetilde{A})$.
Also, a filter ratio $\beta\in [0,1]$ is set for the interference adjacency matrix.
In Eq.~\eqref{eq:interference_weight}, the final interference adjacency matrix $\widetilde{A}^{\circ}$ is calculated, where $i>w$ is used to avoid that future thoughts influence the past, and $ \widetilde{a}'_{j,k} \in \widetilde{A}'$. 
The interference adjacency matrix $\widetilde{A}^{\circ}$ reveals how past events promote future events.
\begin{equation}
\label{eq:interference_weight}
  \widetilde{a}^{\circ}_{j,k} = 
  \begin{cases}
    \widetilde{a}'_{j,k}, &\text{if $i>w$ and}\\
    &\text{$\widetilde{a}'_{j,k} \geq   \beta (max (\widetilde{A}')-min (\widetilde{A}'))+min (\widetilde{A}')$,}\\
	0, &\text{else}
  \end{cases}
\end{equation}

\subsubsection{Graph Convolutional Networks}
To mine the representation of the learned EEG data $x^{\circ}$ aggregated by $\widehat{A}^{\circ}$ and $\widetilde{A}^{\circ}$, the approximate GCN~\cite{KlicperaBG19,wu2020connecting} is applied to learn two representations $\widehat{x}^{\blacksquare}$ and $\widetilde{x}^{\blacksquare}$ and then concatenate them to form the final EEG representation $\overline{x}$.
In the optimization part (see Eq.~\eqref{eq:main_loss}), the $Sigmoid(\cdot)$ function requires a suitable range of inputs to avoid the gradient exploding issue~\cite{9336631}, while the values in $\widetilde{A}^{\circ}$ have a large variance compared with the values in $\widehat{A}^{\circ}$ due to the continuous multiplication in Eq.~\eqref{eq:interference_matrix}.
To prevent the gradient disaster, weights normalization is performed:
\begin{equation}
\label{eq:weights_norm}
A^{\bullet}_{i} = \frac{A^{\circ}_{i}}{sum(A^{\circ}_i)} \text{ with } 1\leq i\leq|\Phi|,
\end{equation}
where $A^{\bullet}_i$ is the $i$-th row of $A^{\bullet}$ that is any one of $\{\widehat{A}^{\bullet},\widetilde{A}^{\bullet} \}$, $A^{\circ}_i$ is the $i$-th row of the corresponding one of $\{\widehat{A}^{\circ},\widetilde{A}^{\circ} \}$, and $sum(A^{\circ}_i)$ is the sum of the elements of the $i$-th row of $A^{\circ}$.
Then, the approximate GCN aggregates the learned EEG data $x^{\circ}$ via Eq.~\eqref{eq:gcn} with the initial $x^{\bullet ,d}=x^{\circ ,d}$ and $d\leq D$ being the depth of GCN.
\begin{equation}
\label{eq:gcn}
    x^{\bullet ,d+1 } = \left [ \xi \Psi(x^{\circ}, (|\Phi|,\Lambda))+ (1-\xi )(A^{\bullet} \Psi(x^{\bullet ,d}, (|\Phi|,\Lambda)) \right ] H^{d},
\end{equation}
where $x^{\bullet ,d+1 }$ being any one of $\{\widehat{x}^{\bullet,d+1},\widetilde{x}^{\bullet,d+1} \}$ is the EEG representation in depth $d+1$, $A^{\bullet}$ is the corresponding one of $\{\widehat{A}^{\bullet},\widetilde{A}^{\bullet} \}$, $H^{d}$ being the corresponding one of $\{\widehat{H}^{d},\widetilde{H}^{d} \}$ is a linear layer at depth $d$, and $\xi $ controls the ratio of $x^{\circ}$.
If there is a future event $\Theta_{m,i}$, then all events $\Theta_{v,w}$ with corresponding weights can influence the event $\Theta_{m,i}$, but the negative influence and the meaningless influence are filtered by ReLU operations. That is, with Eq.~\eqref{eq:gcn}, GCN aggregates the EEG data by utilizing the meaningful matrices to find the high-order influence.
Next, the output of each layer is concatenated to represent the learned representation from two matrices in Eq.~\eqref{eq:cat_rep}.
\begin{equation}
\label{eq:cat_rep}
 x^{\blacksquare}= x^{\bullet ,1 }\oplus x^{\bullet ,2 }\oplus  \cdots \oplus x^{\bullet ,d+1 },
\end{equation}
where $x^{\blacksquare}$ is any of $\{\widehat{x}^{\blacksquare},\widetilde{x}^{\blacksquare} \}$, and $\oplus$ is the concatenation operation.
Eventually, $\widehat{x}^{\blacksquare}$ and $\widetilde{x}^{\blacksquare}$ are concatenated to obtain the final representation $\overline{x}$ in Eq.~\eqref{eq:final_cat}.
\begin{equation}
\label{eq:final_cat}
\overline{x} = Seq_3(Seq_1(\widehat{x}^{\blacksquare}) \oplus Seq_2(\widetilde{x}^{\blacksquare})),
\end{equation}
where $Seq_1(\cdot)=(ReLU(\cdot  H_1)) H_2$, $Seq_2(\cdot)=(ReLU(\cdot  H_3)) H_4$, and $Seq_3(\cdot)=(ReLU(\cdot  H_5)) H_6$ are sequence templates composed of linear layers.
In QUARK, we complete the learning process from the raw data $x$ to the EEG representation $\overline{x}$.

%--------------------------------------------------------------------------
\subsection{Model Training \& Prediction}\label{model_training_and_prediction}
In optimization, the objective function has three parts, including the main loss of Bayesian personalized ranking (BPR)~\cite{bpr}, $ L_{1}$; quantum loss of orthogonality,  $L_{2}$; and continuity loss of EEG data, $L_{3}$.

To apply the BPR algorithm, a sampled item $y_{ \triangleleft}$ with the same label as the predicted EEG data is marked as ``liked'', while $y_{ \triangleright}$ with another different label is marked as ``disliked''.
Hence, the loss of BPR algorithm can be defined in Eq.~\eqref{eq:main_loss}. 
\begin{equation}
\label{eq:main_loss}
L_{1}=\frac{1}{\left | X \right |} \sum_{x\in X}\sum_{y_{ \triangleleft} \in Y_{ \triangleleft} }\sum_{y_{ \triangleright} \in Y_{ \triangleright} }-log(Sigmoid \left ( \overline{x} y_{ \triangleleft} - \overline{x} y_{ \triangleright} \right )),
\end{equation}
where $Y_{ \triangleleft}$ and $Y_{ \triangleright}$ are the sets of liked and disliked items, respectively, and can be trained by any of the existing item representation learning models if the items needs to be pre-processed ({\em e.g.}, images, music, and micro-video).
QUARK is optimized by maximizing the gap between the scores $\overline{x}y_{ \triangleleft}$ and $\overline{x}y_{ \triangleright}$ of the liked and disliked items.

For the orthogonality of quantum theory, the matrix property ({\em i.e.},  $B_{m,i} B_{m,i}^{T}=I$) is considered in Eq.~\eqref{eq:oth_loss}, where $T$ is the transpose,  $I$ is the identity matrix, and $\left \| \cdot \right \|_F$ is the Frobenius norm.
\begin{equation}
\label{eq:oth_loss}
L_{2}=\left \| B_{m,i}B_{m,i}^T-I  \right \|_F.
\end{equation}

Since the event $\Theta_{m,i+1}$ occurs based on the previous event $\Theta_{m,i}$, the loss limits that the learned data should be similar to the next event in Eq.~\eqref{eq:loss_cnt}, where $x_{m,i}^{\circ}$ is the learned data that is estimated as the state in $\Theta_{m,i+1}$, and $x_{m,i+1}^*$ represents the actual state in $\Theta_{m,i+1}$
\begin{equation}
\label{eq:loss_cnt}
L_{3}= \sum_{m< M} \sum_{ i< \mho}-log(Sigmoid({x_{m,i}^{\circ}x_{m,i+1}^*})).
\end{equation}

Finally, the total loss $L$ is computed via Eq.~\eqref{eq:loss_total}, where $\rho$ is the regularization weight~\cite{bisong2019regularization}, and $\Gamma $ contains all the learnable parameters, including all basis vectors in $B$, $\widetilde{H}^{d}|d=1,2,\cdots,D \}, \{ \widehat{H}^{d}|d=1,2,\cdots,D \}, \{H_j|j=1,2,\dots,6 \}$. 
\begin{equation}
\label{eq:loss_total}
L=L_{1}+L_{2}+L_{3}+ \rho  \left \| \Gamma   \right \|_F.
\end{equation}

\begin{table*}[]
\caption{Comparison between baselines and QUARK.}
\label{table:comparison}
\centering
\scalebox{0.8}{
\setlength{\tabcolsep}{4mm}{
\begin{tabular}{cccccccccc}
\toprule
\multirow{2}{*}{Dataset}       & \multirow{2}{*}{Metrics} & (a)          & (b)    & (c)    & (d)    & (e)    & (f)    & (g)             & \multicolumn{1}{l}{Improvement (\%)} \\
                               &                          & Random Guess & MF     & DeepFM & DLRM   & NCF    & BPR    & QUARK           & (g) vs. (f)                          \\
\midrule
\multirow{3}{*}{111\_Normal}   & P@10                     & 0.1530       & 0.1401 & 0.1453 & 0.1472 & 0.1576 & 0.2416 & \textbf{0.3011} & 24.63                                \\
                               & R@10                     & 0.1020       & 0.0930 & 0.0968 & 0.0981 & 0.1051 & 0.1611 & \textbf{0.2007} & 24.58                                \\
                               & F1@10                    & 0.1224       & 0.1120 & 0.1162 & 0.1178 & 0.1261 & 0.1933 & \textbf{0.2409} & 24.62                                \\
\midrule
\multirow{3}{*}{50\_Normal}    & P@10                     & 0.1498       & 0.1427 & 0.1515 & 0.1529 & 0.1541 & 0.2247 & \textbf{0.3070} & 36.63                                \\
                               & R@10                     & 0.0990       & 0.0950 & 0.1011 & 0.1019 & 0.1027 & 0.1498 & \textbf{0.2047} & 36.65                                \\
                               & F1@10                    & 0.1198       & 0.1141 & 0.1213 & 0.1223 & 0.1232 & 0.1797 & \textbf{0.2456} & 36.67                                \\
\midrule
\multirow{3}{*}{25\_Normal}    & P@10                     & 0.1482       & 0.1404 & 0.1416 & 0.1546 & 0.1505 & 0.2566 & \textbf{0.3945} & 53.74                                \\
                               & R@10                     & 0.0980       & 0.0936 & 0.0944 & 0.1031 & 0.1003 & 0.1710 & \textbf{0.2630} & 53.80                                \\
                               & F1@10                    & 0.1186       & 0.1123 & 0.1133 & 0.1237 & 0.1203 & 0.2052 & \textbf{0.3156} & 53.80                                \\
\midrule
\multirow{3}{*}{8\_Normal}     & P@10                     & 0.1541       & 0.1847 & 0.3276 & 0.2651 & 0.2459 & 0.5105 & \textbf{0.6807} & 33.34                                \\
                               & R@10                     & 0.1027       & 0.1231 & 0.2184 & 0.1767 & 0.1639 & 0.3403 & \textbf{0.4538} & 33.35                                \\
                               & F1@10                    & 0.1233       & 0.1478 & 0.2321 & 0.2121 & 0.1967 & 0.4084 & \textbf{0.5445} & 33.33                                \\
\midrule
\multirow{3}{*}{111\_LongTail} & P@10                     & 0.1489       & 0.1378 & 0.1453 & 0.1456 & 0.1488 & 0.1581 & \textbf{0.2034} & 28.65                                \\
                               & R@10                     & 0.0990       & 0.0910 & 0.0969 & 0.0971 & 0.0990 & 0.1054 & \textbf{0.1356} & 28.65                                \\
                               & F1@10                    & 0.1191       & 0.1102 & 0.1163 & 0.1165 & 0.1190 & 0.1265 & \textbf{0.1627} & 28.62                                \\
\midrule
\multirow{3}{*}{50\_LongTail}  & P@10                     & 0.1521       & 0.1405 & 0.1411 & 0.1519 & 0.1423 & 0.1597 & \textbf{0.2261} & 41.58                                \\
                               & R@10                     & 0.1014       & 0.0936 & 0.0941 & 0.1012 & 0.0940 & 0.1064 & \textbf{0.1507} & 41.64                                \\
                               & F1@10                    & 0.1217       & 0.1124 & 0.1129 & 0.1215 & 0.1139 & 0.1277 & \textbf{0.1809} & 41.66                                \\
\midrule
\multirow{3}{*}{25\_LongTail}  & P@10                     & 0.1454       & 0.1417 & 0.1457 & 0.1492 & 0.1494 & 0.1690 & \textbf{0.2601} & 53.91                                \\
                               & R@10                     & 0.0969       & 0.0945 & 0.0971 & 0.0990 & 0.0996 & 0.1126 & \textbf{0.1734} & 54.00                                \\
                               & F1@10                    & 0.1163       & 0.1134 & 0.1165 & 0.1194 & 0.1195 & 0.1352 & \textbf{0.2081} & 53.97                                \\
\midrule
\multirow{3}{*}{8\_LongTail}   & P@10                     & 0.1495       & 0.1645 & 0.1702 & 0.1597 & 0.1499 & 0.2029 & \textbf{0.3971} & 95.71                                \\
                               & R@10                     & 0.0990       & 0.1097 & 0.1134 & 0.1064 & 0.0990 & 0.1353 & \textbf{0.2647} & 95.64                                \\
                               & F1@10                    & 0.1196       & 0.1316 & 0.1361 & 0.1277 & 0.1199 & 0.1623 & \textbf{0.3177} & 95.75          \\          
\bottomrule
\end{tabular}
}}
\end{table*}

\section{Performance Evaluation} \label{Evaluation}
% We validate model performances with ``MindBigData Insight v1.04 Only EEG'' (MBD), which is an open EEG signal database~\cite{mindbigdata} captured in the real world with the stimulus of seeing random images from IMAGENET ILSVRC2013~\cite{ILSVRC15} training dataset and thinking about it.
% We apply MF~\cite{koren2009matrix}, DeepFM~\cite{ijcai2017239}, DLRM~\cite{naumov2019deep}, NCF~\cite{he2017neural}, and BPR~\cite{bpr} as baseline models to validate the performance of QUARK.
% Detailed information on datasets and baseline models are introduced in Section~\ref{datasets} and Section~\ref{baselinemodels}, respectively.
% Experimental settings are described in Section~\ref{Experiment_Settings} with corresponding hyperparameter analysis in Section~\ref{Hyperparameter_Analysis}.

%\subsection{Dataset}\label{datasets}

{\bf Dataset.} %Our proposed model is a general EEG-to-item recommendation framework.
Due to the scarcity of datasets, we apply a visual-EEG-activated dataset to evaluate the performance of our model.
``MindBigData Insight'' (MBD) is an open EEG signal database that contains 569 classes with 14,012 EEG instances~\cite{mindbigdata} captured in the real world with the stimulus of seeing random images from IMAGENET~\cite{ILSVRC15} training dataset and thinking about it.
MBD data records EEG signals when users watch images and contain the users' reflections on different images.
Accordingly, the learned EEG representation can be used to represent what a user watches and thinks, and the images with the category the same as the user's EEG can reflect what the user may want ({\em e.g.}, based on EEG representation of watching a computer,  computers are recommended).

To comprehensively test the performance of our recommendation mechanisms on various granularities of image categories, we adopt the following methods to merge the image categories~\cite{xie2022multi}.
MBD contains two parent categories: the biological categories and the nonbiological categories. 
The child categories in biological categories can be merged if they have the same parent category in biological taxonomic rank~\cite{Taxonomicrank}, and the child categories in nonbiological categories can be merged if Amazon search returns the same parent category when we search for them on Amazon~\cite{Amazoncom}, {\em i.e.}, the parent category indicates a coarse-grained category, while the child category a fine-grained category.
Data distribution is another important factor that influences the performance.
Two kinds of distributions are considered here:
(i) the long tail distribution simulates that some users have similar backgrounds or professions and frequently view the same things;
and (ii) the normal distribution is used to simulate that users have different backgrounds or professions, so they view different things in different scenarios.
The data in MBD follows the long-tail distribution because the dataset publisher collected more EEG instances of viewing animals compared to the other classes.
For the normal distribution, we tune the number of the data instances until the number of instances in all classes (approximately) follows the standard normal distribution~\cite{weisstein2002normal}.
The statistics are summarized in Table~\ref{tables:datasets}.
\begin{table}
\caption{Datasets.}
\label{tables:datasets}
\centering
\scalebox{0.85}{
\setlength{\tabcolsep}{2mm}{
\begin{tabular}{lllll}
\toprule
Dataset & \# of EEG & \# of images & \# of classes & Distribution \\
\midrule
111\_Normal              & 7,048                                                                 & 13,998                                                                   & 111                                                                       & Normal \\
50\_Normal               & 7,048                                                                 & 13,998                                                                   & 50                                                                        & Normal\\
25\_Normal               & 7,048                                                                 & 13,998                                                                   & 25                                                                        & Normal \\
8\_Normal                & 7,048                                                                 & 13,998                                                                   & 8                                                                         & Normal \\
111\_LongTail            & 14,012                                                                & 13,998                                                                   & 111                                                                       & Long tail\\
50\_LongTail             & 14,012                                                                & 13,998                                                                   & 50                                                                        & Long tail\\   
25\_LongTail             & 14,012                                                                & 13,998                                                                   & 25                                                                        & Long tail\\
8\_LongTail              & 14,012                                                                & 13,998                                                                   & 8                                                                         & Long tail\\                   
\bottomrule
\end{tabular}
}}
\end{table}

%To keep the same classes in two distributions, we have to reduce the number of pictures in normal distribution.

% *********************************  baseline ****************************
%\subsection{Baseline Models}\label{baselinemodels}

{\bf Baseline Models.} Since there is no graph relation/information ({\em e.g.}, social, item-item relation, {\em etc}.) in MBD, the state-of-the-art graph-based recommendation model can not be applied on MDB.
Since there is no public framework for EEG-based recommendation, we adopt the following classic but general recommendation models as baseline models. EEG data is processed in the same way.
\textbf{(i) MF}~\cite{koren2009matrix}: MF is a classic user-item recommendation framework based on matrix factorization. \textbf{(ii) DeepFM}~\cite{ijcai2017239}: DeepFM adds extra neural networks based on MF and recommends items by considering the combination of the scores of neural networks and MF. \textbf{(iii) DLRM}~\cite{naumov2019deep}: It is an effective model proposed by Facebook to recommend items via prediction scores. \textbf{(iv) NCF}~\cite{he2017neural}: NCF is a generic recommendation model based on collaborative filtering and neural networks with users and items as input. \textbf{(v) BPR}~\cite{bpr}: BPR is a personalized algorithm that recommends items by maximizing the gap of recommendation scores between liked items and disliked items. (vi) {\bf Random Guess}: It randomly recommends images to investigate whether the learning-based models can perform better.

%-------------------------------------------------------------------------
%\subsection{Experiment Settings}\label{Experiment_Settings}
{\bf Experiment Settings.} All images in MBD are trained by the state-of-the-art model, ViTAE2~\cite{zhang2022vitaev2}, to obtain the image representation for QUARK and baselines. 
The training rate for all experiments is 0.85, where 85\% of MBD data is used for training and the remaining 15\% is used for testing.
According to MBD, the number of electrodes is set as $M=5$, and the length of time step is set as $N=360$.
% which is ranked as top 2 in the benchmark of the IMAGENET dataset.
To better train QUARK, the learnable parameters in $\Gamma$ (see Eq.~\eqref{eq:loss_total}) are initialized through Xavier distribution~\cite{glorot2010understanding} and optimized by mini-batch Adaptive Moment Estimation (Adam)~\cite{adam}. 
We have conducted intensive experiments with different hyperparameter settings to find the appropriate parameter values.
Specifically, the learning rate is set as 1e-4, mini-batch size is set as 16, the regularization weight in Eq.~\eqref{eq:loss_total} is set as $\rho=1e-4$, the set of hyperparameters $\{\Lambda =15, \Delta =25, |B_{m,i}|=15, c=2, \alpha = 0.8, \beta = 0.4, d=5,  \xi= 0.3 \}$ is used for the normal distribution, and the set of hyperparameters $\{ \Lambda =10, \Delta =20, |B_{m,i}|=10, c=2, \alpha = 0.9, \beta = 0.7, d=5, \xi= 0.3 \}$ is applied for the long tail distribution (please see supplement materials [\ref{outside_link}] Section 2 for more hyperparameter analysis). 
We follow the recommendation settings in~\cite{example3} to randomly sample 100 images as candidates, including 15 images with the same class of the trained EEG and 85 images with random classes except the class of the trained EEG.
The recommendation performance is measured by the averaged Precision@k score that demonstrates the precision of the recommendation~\cite{fayyaz2020recommendation}, the averaged Recall@k score that indicates how many correct images are recommended from the sampled correct images~\cite{fayyaz2020recommendation}, and the averaged F1@k score that is the mixture of P@k and R@k and provides guidance if the performance cannot be distinguished only from Precision@k and Recall@k~\cite{fayyaz2020recommendation}. 
In our experiments, the number of recommended images is $k=10$.

%==================================================================

% ***********  Evaluation- Comparison *****************
%\subsection{QUARK {\em v.s.} Baselines} \label{Performance_Comparison}

{\bf QUARK {\em v.s.} Baselines.} The comparison results are summarized in Table~\ref{table:comparison}, where the improvement is the performance increase in QUARK compared with the baseline.
From Table~\ref{table:comparison}, three main results are concluded. 
Firstly, QUARK outperforms all baselines under all experiment scenarios, due to the following reasons:
% Compared to BPR that has the best performance among the baselines, p@k is improved 27.69\%, 33.57\%, 28.64\%, and 95.71\% on dataset 111\_Normal, 8\_Normal, 111\_LongTail, and 8\_LongTail, respectively.
(1) the sliding window-based segmented EEG data can represent different thoughts as the bionic in the temporal domain; (2) quantum space is employed to extract the various thinking factors; (3) two graph adjacency matrices simulate the way that a person thinks; and (4) the GCN aggregates the multi-depth information as a user's preference. 
Secondly, with the number of classes increasing, the performance of all models is decreased because the EEG-based recommendation task becomes more difficult.
When the number of classes is larger than 8, the classic recommendation models (b)-(e) perform equally to the random guess model (a). 
BPR outperforms random guess model because it can build a gap to distinguish liked and disliked images as QUARK does, but its performance is still worse than that of QUARK. 
Thirdly, compared with normal distribution, it is harder to perform recommendations with images following long tail distribution. This is because the models are trained on a large number of data from a few classes, which makes the model parameters trend to recommend the images of those few classes for any incoming EEG data.
Through the comparison, higher performance improvements under long tail distribution are shown by comparing QUARK with BPR, implying that the novel design of QUARK ensures it works well on difficult recommendation tasks.
%These results in Table~\ref{table:comparison} emphasize the validity of QUARK by using well-designed components such as sliding windows, quantum spaces, quantum theory, adjacency masks, and GCNs.

\begin{figure}[ht]
\centering
    \subfigure[]{\includegraphics[width=0.09\textwidth]{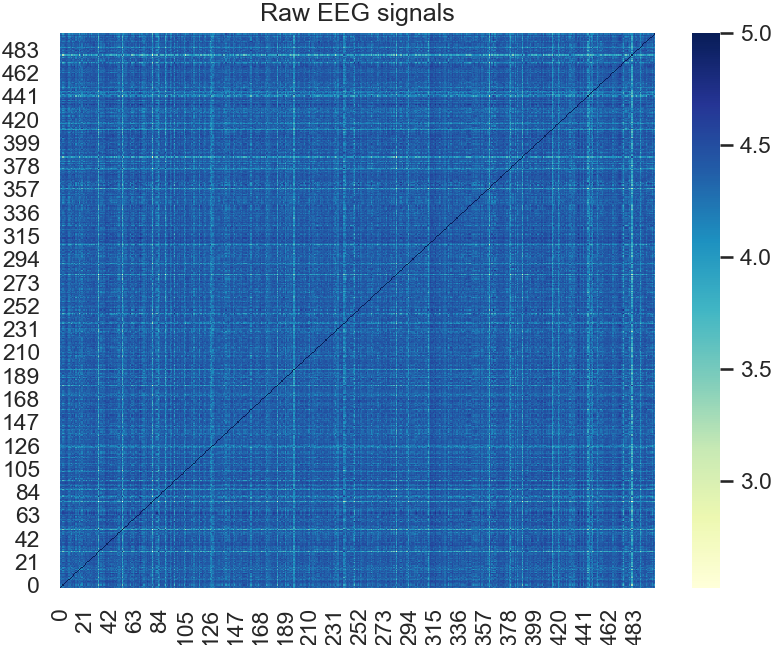}\label{fig:raw_eeg}}
    \subfigure[]{\includegraphics[width=0.09\textwidth]{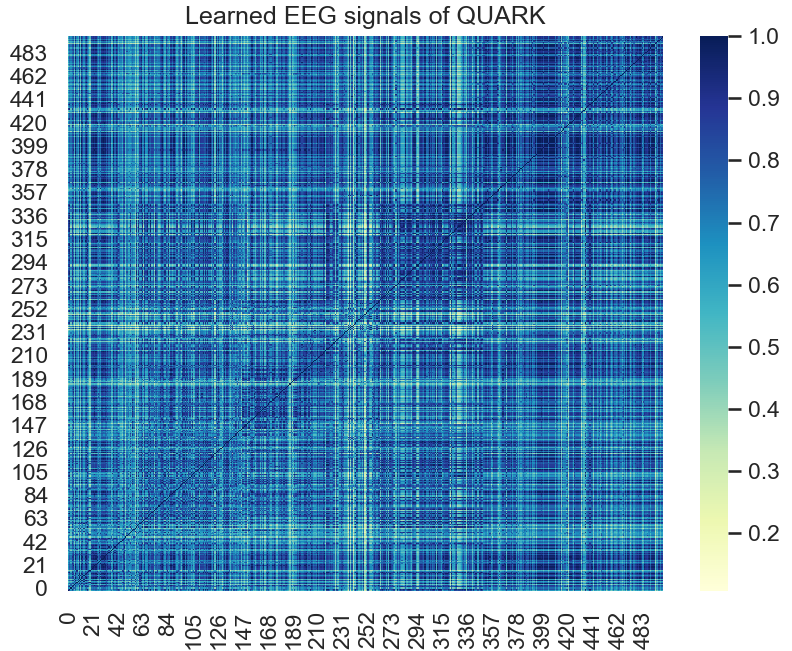}\label{fig:learned_eeg_quark}} 
    \subfigure[]{\includegraphics[width=0.09\textwidth]{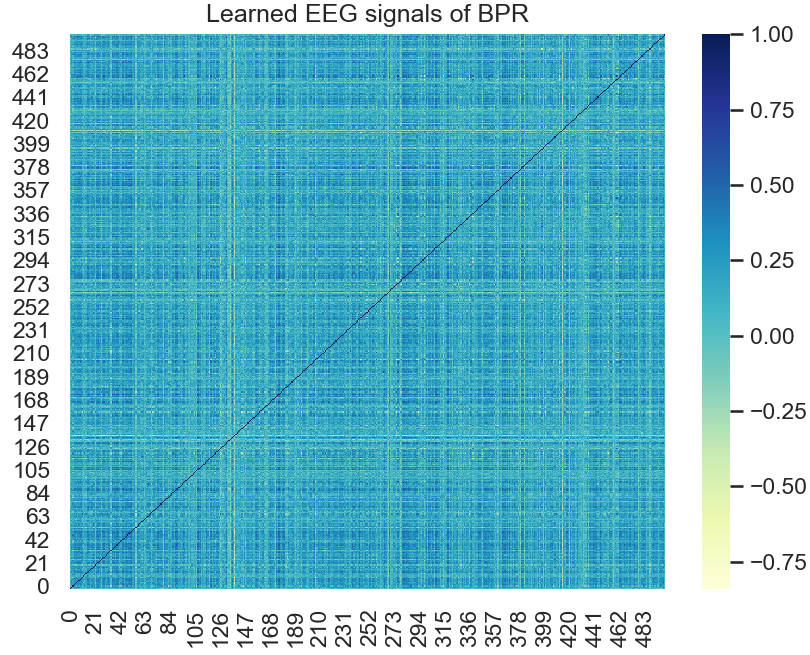}\label{fig:learned_eeg_bpr}} 
    \subfigure[]{\includegraphics[width=0.09\textwidth]{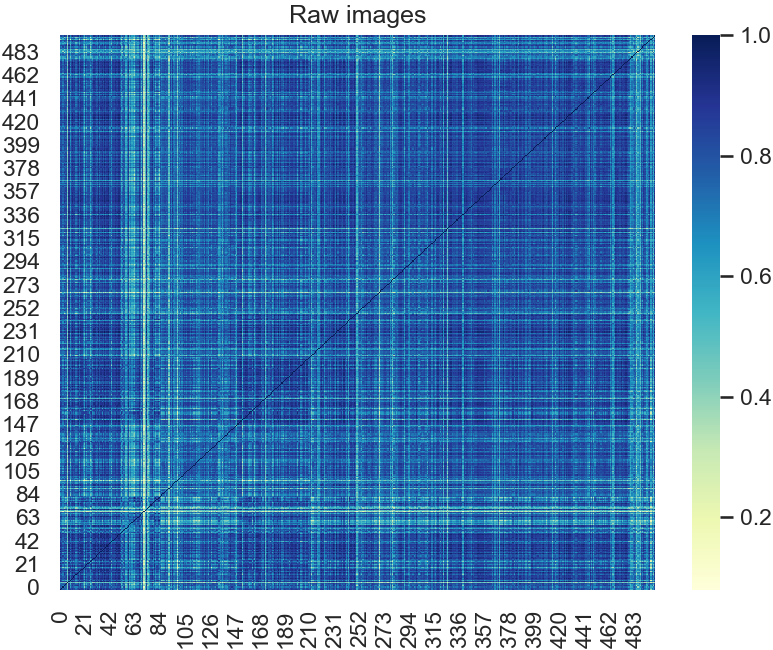}\label{fig:raw_images}}
    \subfigure[]{\includegraphics[width=0.09\textwidth]{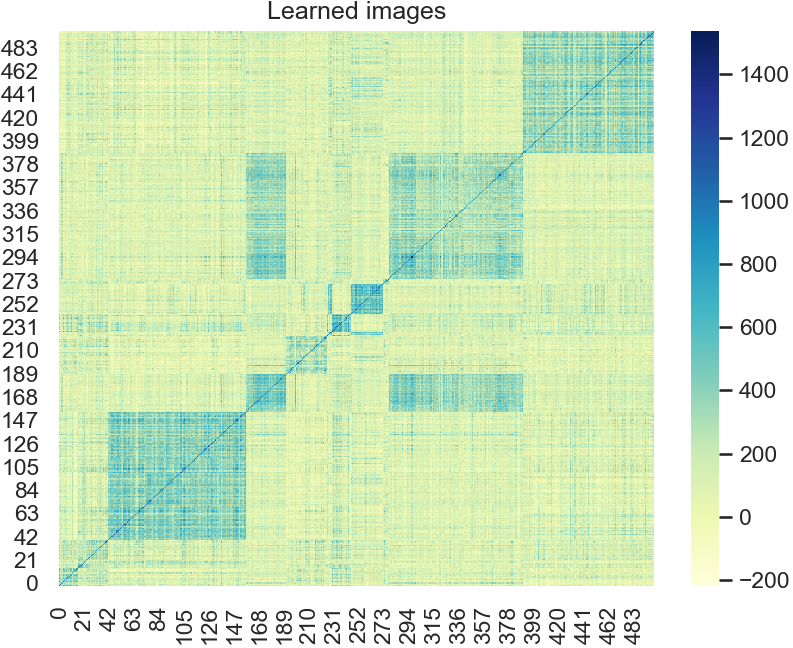}\label{fig:learned_images}}
\caption{\small Representation visualization. (a) Similarity of raw EEG signals, (b) Similarity of representation learned by QUARK, (c) Similarity of representation learned by BPR, (d) Similarity of raw image data, and (e) Similarity of learned image representation.}
\label{fig:representations}
\end{figure}

{\bf Data \& Representation.}
In Figure~\ref{fig:representations}, each value is the dot similarity between two data instances in MBD.
The similarity is expected to be a higher value (dark blue color) if two data instances are from the same class, and the data from the same class form a cluster ({\em i.e.}, in Figure~\ref{fig:learned_images}, blue cluster represents the same class, while yellow means different classes).
Figure~\ref{fig:raw_images} shows the similarities of raw image information, and Figure~\ref{fig:learned_images} shows the similarities of learned image information using the image recognition model, ViTAE2.
The representations learned by the best baseline model, BPR, in Figure~\ref{fig:learned_eeg_bpr} are similar to the raw EEG in Figure~\ref{fig:raw_eeg}, which means the baselines hardly learn good representations and clear clusters.
While, the representations learned by QUARK in Figure~\ref{fig:learned_eeg_quark} show relatively clear clusters, which illustrates QUARK can learn high-quality representations visualized in Figure~\ref{fig:learned_eeg_quark} and thus can present excellent performance improvement in Table~\ref{table:comparison}.
It explains why the baseline models perform closely to the random guess. 

%\subsection{Ablation Study}

\begin{table*}[]
\caption{Ablation studies of QUARK.}
\label{table:ablation}
\centering
\scalebox{0.85}{
\setlength{\tabcolsep}{3mm}{
\begin{tabular}{cccccccccc}
\toprule
\multirow{2}{*}{Dataset}       & \multirow{2}{*}{Metrics} & (a)     & (b)    & (c)              & (d)          & (e)      & (f)              & (g)          & (h)        \\
                               &                          & $\neg$GCN  & $\neg$QM &$\neg$Interference & $\neg$Continuity & $\neg$ $(i>w)$ & $\neg$Continuity loss & $\neg$QM loss & QUARK \\
\midrule
\multirow{3}{*}{111\_Normal}   & P@10                     & 0.2734  & 0.2789 & 0.2820            & 0.2836       & 0.2859   & 0.2933           & 0.2970        & \textbf{0.3011}     \\
                               & R@10                     & 0.1823  & 0.1859 & 0.1880            & 0.1891       & 0.1906   & 0.1955           & 0.1980        & \textbf{0.2007}     \\
                               & F1@10                    & 0.2187  & 0.2231 & 0.2256           & 0.2269       & 0.2287   & 0.2347           & 0.2376       & \textbf{0.2409}    \\

\midrule
\multirow{3}{*}{111\_LongTail} & P@10                     & 0.1705  & 0.1844 & 0.1862           & 0.1931       & 0.1989   & 0.2021           & 0.2024       & \textbf{0.2034}     \\
                               & R@10                     & 0.1137  & 0.1229 & 0.1241           & 0.1287       & 0.1326   & 0.1347           & 0.1349       & \textbf{0.1356}     \\
                               & F1@10                    & 0.1364  & 0.1475 & 0.1490            & 0.1545       & 0.1591   & 0.1617           & 0.1619       & \textbf{0.1627}     \\
\bottomrule
\end{tabular}
}}
\end{table*}
{\bf Ablation Study.}
In the ablation study, the variants of QUARK are compared to understand the contributions of each component developed in QUARK, and the results of P@10, R@10, and F1@10 are presented in Table~\ref{table:ablation} by testing 111\_Normal and 111\_LongTail. 
These variants include: 
(a) \textbf{$\neg$GCN}, where GCNs are removed, and the adjacency matrices are fed into linear layers instead;
(b) \textbf{$\neg$QM}, where quantum space is removed, and two random adjacency matrices are used; 
(c) \textbf{$\neg$Interference}, where the interference adjacency matrix is removed; 
(d) \textbf{$\neg$Continuity}, where the continuity adjacency matrix is removed; 
(e) \textbf{$\neg(i>w)$}, where the requirement of $i>w$ in Eq.~\eqref{eq:relation_retention_percentage} and Eq.~\eqref{eq:interference_weight} is removed; 
(f) \textbf{$\neg$Continuity loss}, where the continuity loss $L_3$ is removed from the objective function; 
(g) \textbf{$\neg$QM loss}, where the orthogonal loss $L_2$ is removed from the objective function.
%(h) \textbf{$\neg$FQUARK}, where QUARK contains all components.
%
The results in Table~\ref{table:ablation} show that all the developed components have significant contributions to the recommendation performance. 
Especially, GCN is the most important component, because it generates aggregated information that contains the similarity and interference among different thoughts, which is crucial to identify a person's thoughts.
Quantum space is secondly important, as it contributes not only to the generation of interference and continuity matrices but also to the decomposition of EEG signal via collapse. 
Besides, the importance of the interference matrix is more significant than that of the continuity matrix because the interference matrix reveals how past thoughts influence future thoughts, which is a bionic process.
Moreover, the requirement of $i>w$ provides contributions by avoiding the influence of future events on past events.
Last but not least, multiple optimization goals in the objective function can help QUARK better recommend items.

% %==================================================================

%\subsection{Feeling and Style Detection}\label{FnSD}
%
\begin{figure*}
\centering
    \subfigure[]{\includegraphics[width=0.18\textwidth]{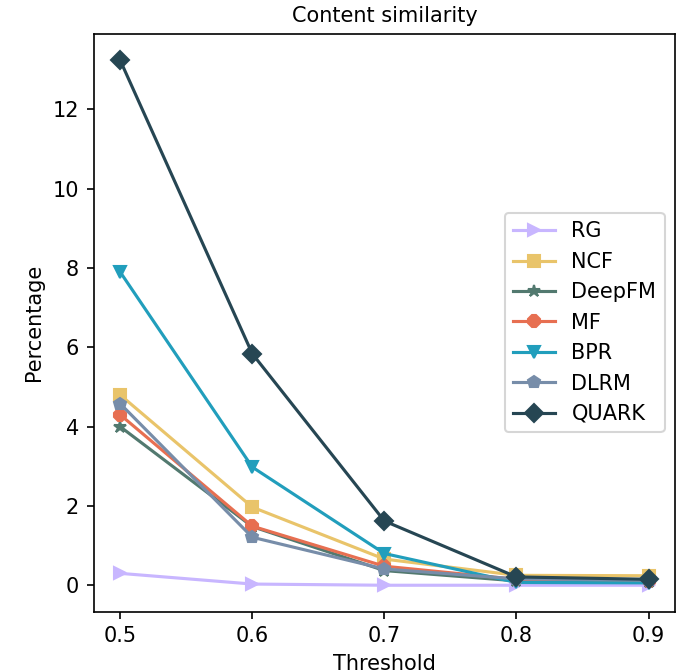}\label{fig:content}}\quad 
    \subfigure[]{\includegraphics[width=0.18\textwidth]{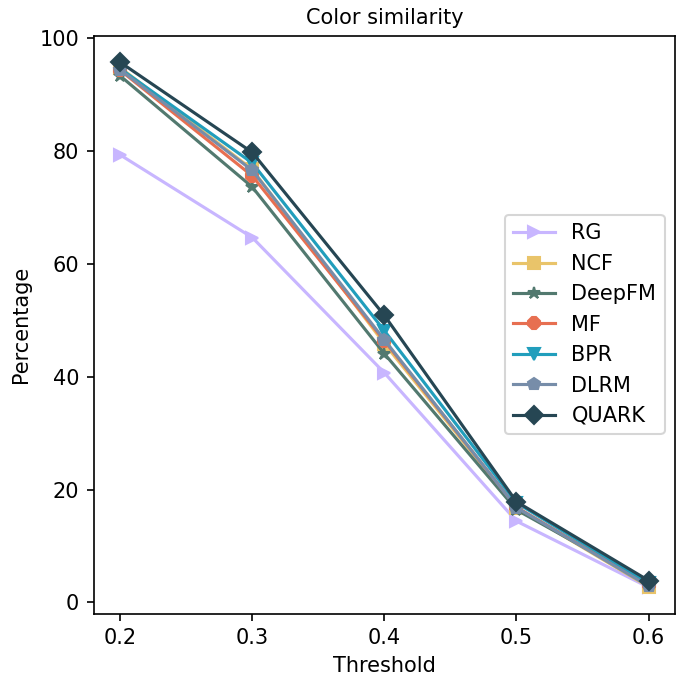}\label{fig:color}}\quad
    \subfigure[]{\includegraphics[width=0.18\textwidth]{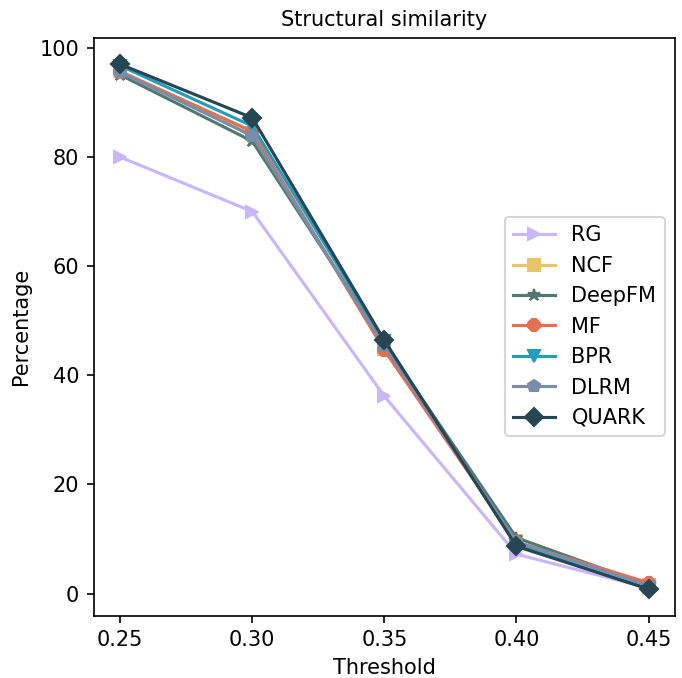}\label{fig:strc}}\quad 
    \subfigure[]{\includegraphics[width=0.18\textwidth]{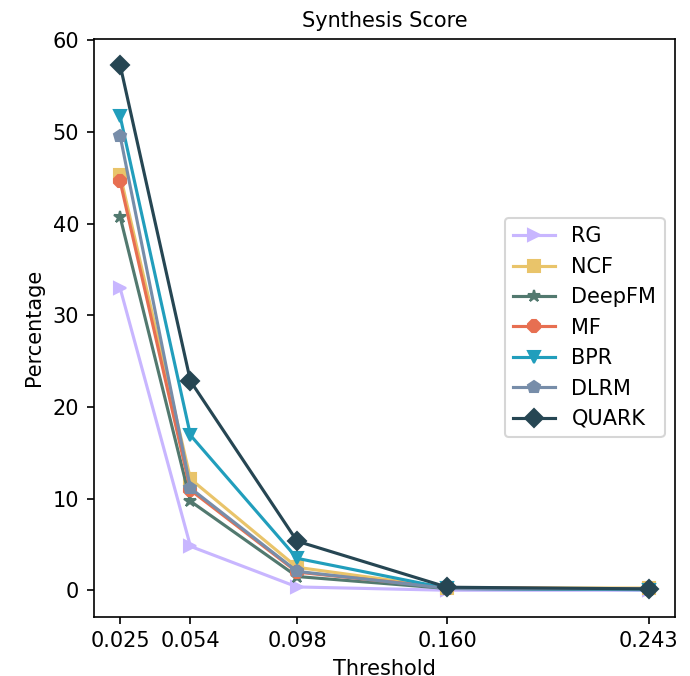}\label{fig:syn}}\quad
    \subfigure[]{\includegraphics[width=0.18\textwidth]{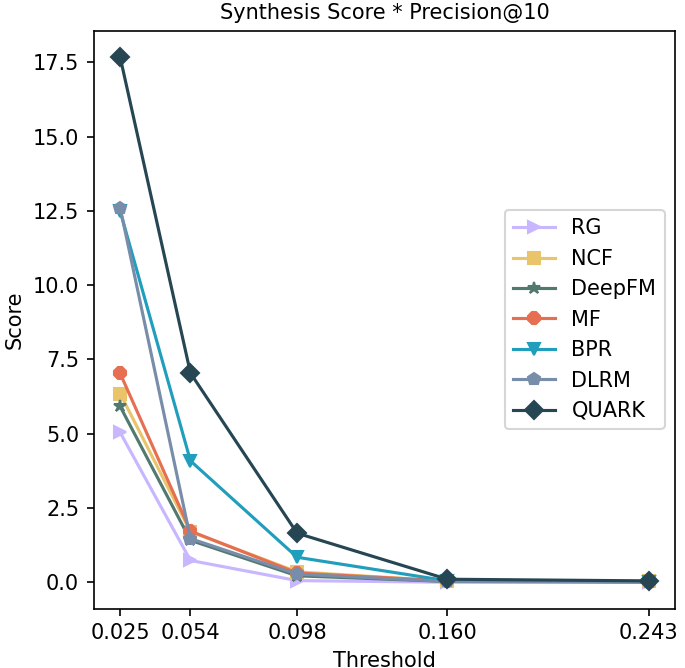}\label{fig:syn_pre}}
\caption{\small The feeling and style of the recommended images compared with the original EEG images. (a) Content similarity. (b) Color similarity. (c) Structural similarity. (d) Synthesis score. (e) Synthesis score with precision@10.}
\label{fig:style_n_feeling}
\end{figure*}

{\bf Feeling \& Style Detection.} EEG signals can reflect users' feelings and personal tastes, which are the crucial factors of the recommendation tasks. 
We design the feeling and style detection to validate whether our model can recognize the user's tastes and focus.
Typically, a person's thoughts can be reflected via the feelings and perceived styles of what the person is viewing, which can be detected from different aspects (please see supplement materials [\ref{outside_link}] Section 3 for following metrics.): (1) content similarity that is the dot similarity of two learned image representations, (2) color similarity that is the difference of the color histogram distribution of two images~\cite{novak1992anatomy}, and (3) structural similarity that is the ratio of the overlapped area of two edge-detected images to the area of one image~\cite{vincent2009descriptive}.
The synthesis score is the multiplication of content, color, and structure similarity, and the mixed score is the multiplication of the synthesis score and Precision@10.
In our experiments, the content score is used to check feeling, and the color and structural scores are used to study style.
Figure~\ref{fig:content} to Figure~\ref{fig:syn_pre} presents the percentage of images counted from 10 recommended images, where a recommended image is counted if the metric score of the recommended image is greater than the corresponding threshold.
Our QUARK is better at recommending images with similar feelings and similar perceived styles, compared with baselines.
The superiority of QUARK comes from quantum space that decomposes EEG signals based on the number of selected basis vectors, each of which is a pure state of a thinking factor containing the feeling or the perceived style, such as the mentioned association, imagination, and {\em etc}. 
As a result, QUARK is not only a good recommendation model with high recommendation precision but also a good feeling and style detector for EEG, which is extremely important for smart personalized recommendations.
On the other hand, recommending the pictures of the same class is regarded as a correct recommendation.
However, pictures have differences even in the same class.
The feeling and style detection can explain that QUARK knows what users see and recommends similar items of the same class. 
%so this work is a recommendation work understanding users' preferences and content and making recommendations instead of a classification task. 

%=========================================================
%\section{Case Study}\label{casestudy}
\begin{figure*}
\centering
    {\includegraphics[width=0.45\textwidth]{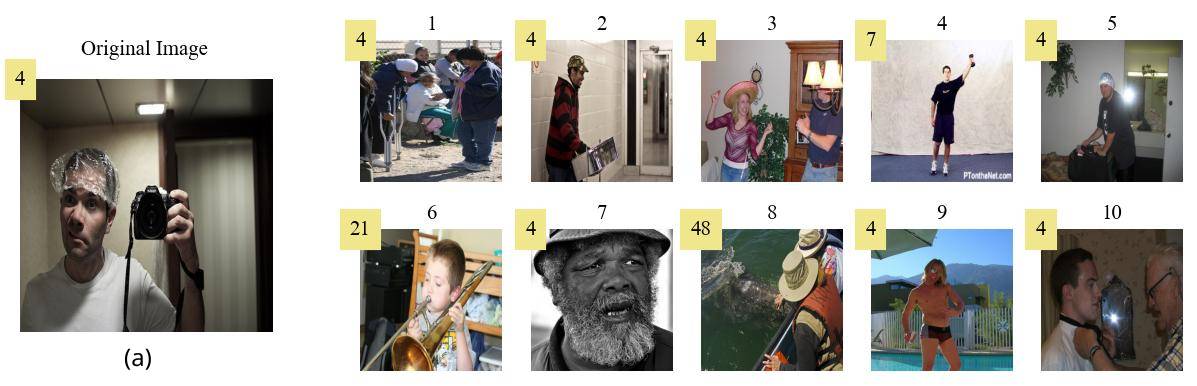}\label{fig:ideal2}}
    {\includegraphics[width=0.45\textwidth]{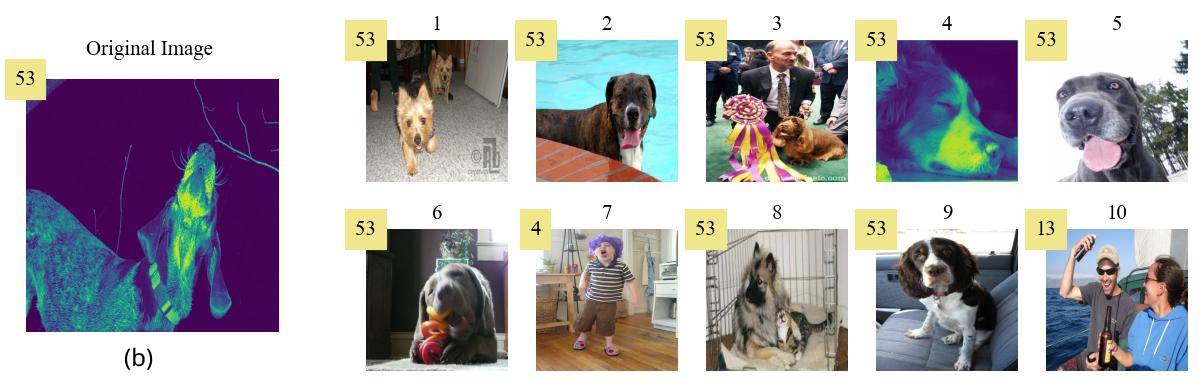}\label{fig:ideal3}}
    {\includegraphics[width=0.45\textwidth]{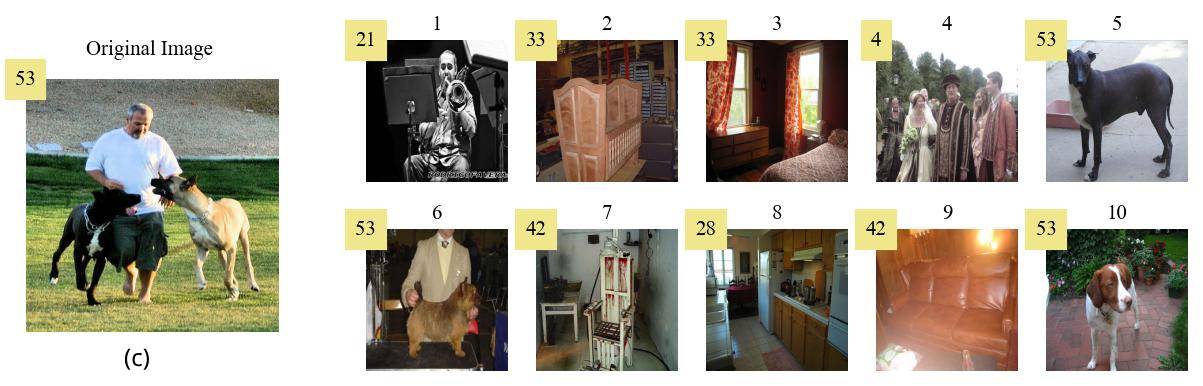}\label{fig:ideal4}}
    {\includegraphics[width=0.45\textwidth]{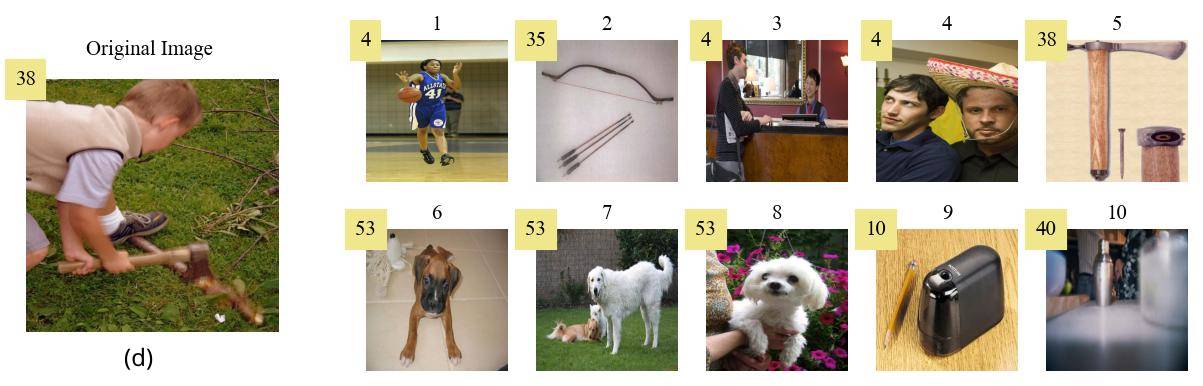}\label{fig:bad1}}
\caption{Case studies.}
\label{fig:Casestudy3}
\end{figure*}

{\bf Case Study}. In this part, some cases are presented in Figure~\ref{fig:Casestudy3} to visualize the recommendation performance of QUARK, where the numbers in the yellow squares represent the classes of images, and the numbers above the images represent the IDs of the recommendations.
\textbf{(1) Color analysis:} a user is looking at the image that a person stands at front of a mirror with the black filter in Figure~\ref{fig:Casestudy3} group (a). Most of the images ({\em i.e.} images 1,2,5,7,8, and 10) recommended by QUARK are with a black filter. 
Another example is that a user is thinking about the original image in Figure~\ref{fig:Casestudy3} group (b), and some of the recommended images are with the blue filter, such as images 2, 4, and 10. These cases demonstrate that QUARK can extract the feelings and styles of the color that the original conveys.
\textbf{(2) Structural analysis:} QUARK can understand the structural information of an image from a person's thoughts, as shown in Figure~\ref{fig:Casestudy3} group (b), where image 2 with nearly the same structural information as the original image is recommended. 
Moreover, from the viewpoint of the edge detection algorithm, we can claim that most of the recommended dogs are in a ``corner'', such as images 1, 2, 4, 5, 6.
Another example is Figure~\ref{fig:Casestudy3} group (c), where the original image contains the structural information of three parts, {\em i.e.}, ``dog-person-dog''. 
For the recommendation, image 1 shows ``device-person-device'', image 3 presents ``window-window-bed'', image 4 displays ``person-person-person'', and image 9 looks like ``sofa-sofa-sofa''. 
\textbf{(3) Wrong recommendation analysis:} in Figure~\ref{fig:Casestudy3} group (d), the precision@10 score is 0.1 as there is one recommended image with the correct label.
The recommendation seems not good, but actually, the content, color, and structural information of EEG can still be understood.
In Figure~\ref{fig:Casestudy3} group (d), the original image shows that a boy is cutting a stick with an ax, which is only labeled as class 38 which is the class of ax.
Some images about humans are recommended, but these images are not counted as correct recommendations.
This incorrect recommendation is due to the single label of the dataset.
Therefore, QUARK can be improved if a multi-label dataset is used.

%==================================================================
% ******************  Conclusion ***********
\section{Discussion \& Conclusion}\label{Conclusion}
To explore the next-generation recommendation, this paper proposes a novel
Quantum Cognition-Inspired EEG-based Recommendation by using Graph Neural Network (QUARK).
Our QUARK model has tightly coupled components, including sliding windows to segment thoughts, quantum space to mine latent representations, quantum space to generate continuity and interference adjacency matrices, and graph convolutional networks to aggregate learned latent EEG information. 
The experiments can confirm the advantages of our QUARK compared with the classic recommendation models on a real-world dataset with extensive settings. 
In the adopted EEG dataset, MBD, each EEG instance and its corresponding image have one label.
In fact, the images contain multiple objects, so the viewer might be distracted by unlabeled object(s), which may reduce the accuracy of preference prediction and the precision of image recommendation.
We believe that QUARK can be better if a multi-label dataset is used.
%From another perspective, if this is a online shopping recommendation, all images are super clear with one object, our QUARK is also able to show better performance because irrelevant object(s) are removed, and EEG signals do not contain the information of other object(s).
On the other hand, since a user's preference can be affected by complicated factors, the recommendation performance can be further enhanced if multi-source data are available for use, such as EEG, user historical interests, and recent hot spots.

For the generality of QUARK, the experiments validate QUARK performance well when the EEG signal and items are in the same domain ({\em i.e.} see images and recommend images in our experiments). 
While, we believe QUARK can efficiently work in cross-domain.
For example, QUARK learns the thoughts when a user watches a romance movie and recommends sweet music or micro-videos.
As we do not have such a dataset, this task is left as our future work.

Through this paper, one can see that EEG-based recommendation has great potential in high-tech fields. 
This is the first detailed EEG-based recommendation model to conduct a study in such an area, and more efforts will be devoted to our future research.

\begin{acks}
This work is supported by U.S. National Science Foundation grants, No. 2343619, 2244219, 2315596, 2146497, and 2416872.
\end{acks}

\newpage

\balance
\bibliographystyle{ACM-Reference-Format}
\bibliography{reference}  

\appendix
\section{Interference between Two Events} \label{appendix_interference}
An orthogonal matrix $B_{m,i}$ has the property $B_{m,i}B_{m,i}^{\dagger}=  \left [ | b_1\rangle ,| b_2\rangle ,\cdots,|b_{\left | B_{m,i} \right |}\rangle   \right ] \left [ | b_1\rangle ,| b_2\rangle ,\cdots,|b_{\left | B_{m,i} \right |}\rangle   \right ]^T = \sum_{j\in |B_{m,i}|}{|b_j\rangle \langle b_j|} = I = (o_{\Theta_{m,1}}+o_{\bar{\Theta}_{m,1}})$ with conjugate transpose $\dagger$, identity matrix $I$ and $b_j\in \mathbb{R}^{\Lambda\times 1}$.
Therefore, the probability $p(\Theta_{m,2})$ can be presented as Eq.~\eqref{eq:pif_1} 
\begin{subequations}
\begin{align}
&p(\Theta_{m,2})=\left \| o_{\Theta_{m,2}}|x\rangle \right \|^2 = \left \| o_{\Theta_{m,2}}I|x\rangle \right \|^2 = \left \| o_{\Theta_{m,2}}(o_{\Theta_{m,1}}+o_{\bar{\Theta}_{m,1}})|x\rangle \right \|^2 \label{eq:pif_1}\\
&=  \langle x |(o_{\Theta_{m,1}}+o_{\bar{\Theta}_{m,1}})o_{\Theta_{m,2}}o_{\Theta_{m,2}}(o_{\Theta_{m,1}}+o_{\bar{\Theta}_{m,1}})|x\rangle \notag \\ %\label{eq:pif_2}\\
&=\langle x |o_{\Theta_{m,1}}o_{\Theta_{m,2}}o_{\Theta_{m,2}}o_{\Theta_{m,1}}|x\rangle + \langle x |o_{\bar{\Theta}_{m,1}}o_{\Theta_{m,2}}o_{\Theta_{m,2}}o_{\bar{\Theta}_{m,1}}|x\rangle  \notag \\
&\langle x |o_{\Theta_{m,1}}o_{\Theta_{m,2}}o_{\Theta_{m,2}}o_{\bar{\Theta}_{m,1}}|x\rangle + \langle x |o_{\bar{\Theta}_{m,1}}o_{\Theta_{m,2}}o_{\Theta_{m,2}}o_{\Theta_{m,1}}|x\rangle \notag\\
&=\left \|  o_{\Theta_{m,2}}o_{\Theta_{m,1}}|x\rangle \right \|^2 + 
\left \| o_{\Theta_{m,2}}o_{\bar{\Theta}_{m,1}}|x\rangle  \right \|^2  \notag  \\
& + \langle x |o_{\Theta_{m,1}}o_{\Theta_{m,2}}o_{\Theta_{m,2}}o_{\bar{\Theta}_{m,1}}|x\rangle  + \langle x |o_{\bar{\Theta}_{m,1}}o_{\Theta_{m,2}}o_{\Theta_{m,2}}o_{\Theta_{m,1}}|x\rangle  \label{eq:pif_3} \\
& =  p(\Theta_{m,2})p(\Theta_{m,1})+p(\Theta_{m,2})p(\bar{\Theta}_{m,1}) \label{eq:pif_4} \\
& + 2 \times \langle x| o_{\bar{\Theta}_{m,1}} o_{\Theta_{m,2}}o_{\Theta_{m,2}} o_{\Theta_{m,1}}|x\rangle  \label{eq:pif_5}\\
&=p'(\Theta_{m,2}) + 2 \times \tau \langle x| o_{\bar{\Theta}_{m,1}} o_{\Theta_{m,2}}o_{\Theta_{m,2}} o_{\Theta_{m,1}}|x\rangle \label{eq:pif_6}
\end{align}
\end{subequations}
In Eq.~\eqref{eq:pif_4}, $ p(\Theta_{m,2})p(\Theta_{m,1})+p(\Theta_{m,2})p(\bar{\Theta}_{m,1}) $ that equals $p'(\Theta_{m,2})$ in Eq.~\eqref{eq:pif_6} is the probability that represents $\left \|  o_{\Theta_{m,2}}o_{\Theta_{m,1}}|x\rangle \right \|^2 + \left \| o_{\Theta_{m,2}}o_{\bar{\Theta}_{m,1}}|x\rangle  \right \|^2$ by assuming that events $\Theta_{m,1}$ and $\Theta_{m,2}$ are independent just as used in CPT, while the rest term in Eq.~\eqref{eq:pif_5} is an extra value that breaks the assumption of independence, which is reformulated from $ \langle x |o_{\Theta_{m,1}}o_{\Theta_{m,2}}o_{\Theta_{m,2}}o_{\bar{\Theta}_{m,1}}|x\rangle +
 \langle x |o_{\bar{\Theta}_{m,1}}o_{\Theta_{m,2}}o_{\Theta_{m,2}}o_{\Theta_{m,1}}|x\rangle$ because $o_{\Theta_{m,1}}o_{\Theta_{m,2}}o_{\Theta_{m,2}}o_{\bar{\Theta}_{m,1}}$ is the conjugation of $o_{\bar{\Theta}_{m,1}}o_{\Theta_{m,2}}o_{\Theta_{m,2}}o_{\Theta_{m,1}}$ in Eq.~\eqref{eq:pif_3}.
Therefore, a parameter $\tau=0$ is involved in maintaining independence in Eq.~\eqref{eq:pif_6} or $\tau= \pm 1$ reveals the interference between two events.
In EEG processing, past thoughts definitely interfere with future thoughts, so $\tau$ is set to 1.
\section{Hyperparameter Analysis}~\label{Hyperparameter_Analysis}
In this section, we discuss the selection and settings of hyperparameters in our experiments. 
As the datasets exhibit precision@10 disparities, plotting the changes of a hyperparameter together may result in plots resembling mere horizontal lines.
To address this issue, we adopt the lowest precision of each dataset in a plot as 0. The remaining precisions are adjusted by subtracting the corresponding lowest precision from each dataset plotted.
This approach allows for clearer visualization of hyperparameter effects. Refer to Figure~\ref{fig:hyperparameters} for a visual representation of these hyperparameter impacts.

\subsubsection{Impacts of window size $\Lambda$.}
The window size represents the amount of information ({\em i.e.},EEG data in this paper) contained in a user's thoughts. 
In Figure~\ref{fig:hyperparameter:win_size}, there are some suitable window sizes -- 10 and 15 for long-tail distribution and normal distribution, respectively.
When the window size is too small, the amount of information in the user's thought is insufficient for learning.
With the precision difference increasing from the window size 5 to 10, more EEG data are contained in a window to represent a latent thought and contain enough data as a state to learn user preference. 
When the window size continues to increase to 35, too much EEG data is contained to reflect more complicated thoughts, which makes QUARK hard to learn what thinking factors are in the user's thoughts.

\subsubsection{Impacts of sliding size $\Delta$.}
The sliding size represents the sample rate of the EEG.
If the sliding size is smaller than the suitable window size, some overlapped EEG data are sampled as inputs and redundant EEG information impairs learning.
On the contrary, if the sliding size is bigger than the suitable window size, some EEG data fragments are not used as inputs to learn the user's thoughts and needs.
In Figure~\ref{fig:hyperparameter:sliding_size}, the suitable sliding sizes are 20 and 25 for long-tail distribution and normal distribution, respectively, which are both bigger than the window sizes. 
QUARK is not that sensitive to the sliding size as 20 to 25 is a good selection for eight datasets with different distributions and different recommendation granularities.
This is because QUARK self-adaptively learns that not all the EEG signals contribute to learning the user's thoughts and needs.

\subsubsection{Impacts of the number of the basis vectors $|B_{m,i}|$.}
The number of basis vectors, $|B_{m,i}|$, shows how many latent thinking factors are contained in a user's thought.
In Figure~\ref{fig:hyperparameter:number_of_b_m_i}, the suitable number of basis vectors are 10 and 15 for long-tail distribution and normal distribution, respectively.
The number of basis vectors less than the suitable values can not provide enough choices for thought to collapse, while too many basis vectors may make thought collapse to wrong basis vectors.

\subsubsection{Impacts of the number $c$ of the selected basis vectors.}
The number of the selected basis vectors, $c$, reveals how many latent factors each thought finally decides to collapse.  
The suitable value of $c$ is 2 for both distributions in Figure~\ref{fig:hyperparameter:number_of_selected_b_m_i}. 
One thinking factor may not be able to describe a user's feelings, while the number of thinking factors more than the suitable value may make EEG representation over-smooth technically when computing a new state.

\subsubsection{Filter ratios $\alpha$ and $\beta$.}
\begin{figure*}
\centering
  \subfigure[Impact of window size $\Lambda$.]{
    \includegraphics[width=0.23\textwidth]{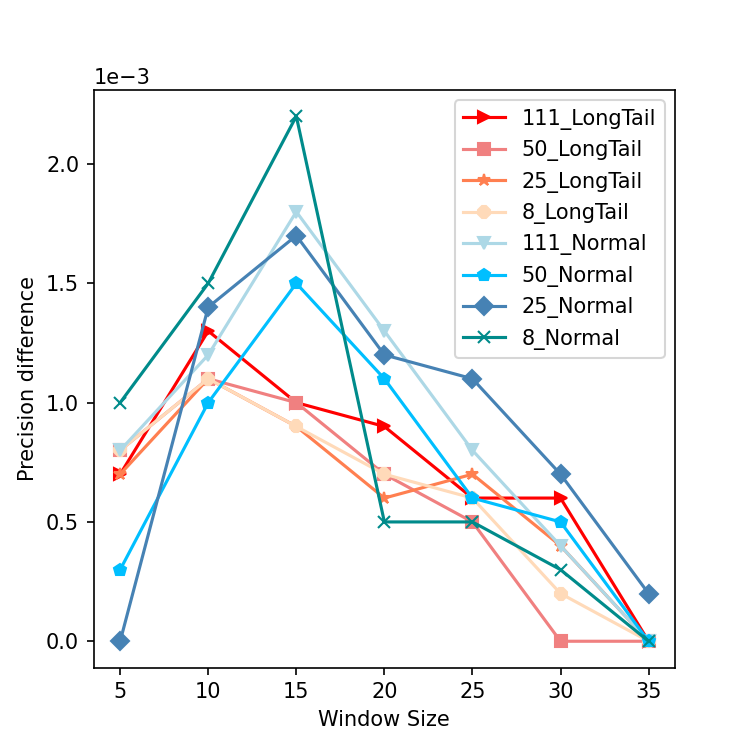}
    \label{fig:hyperparameter:win_size}
    }
    \subfigure[Impact of sliding size $\Delta$.]{
    \includegraphics[width=0.23\textwidth]{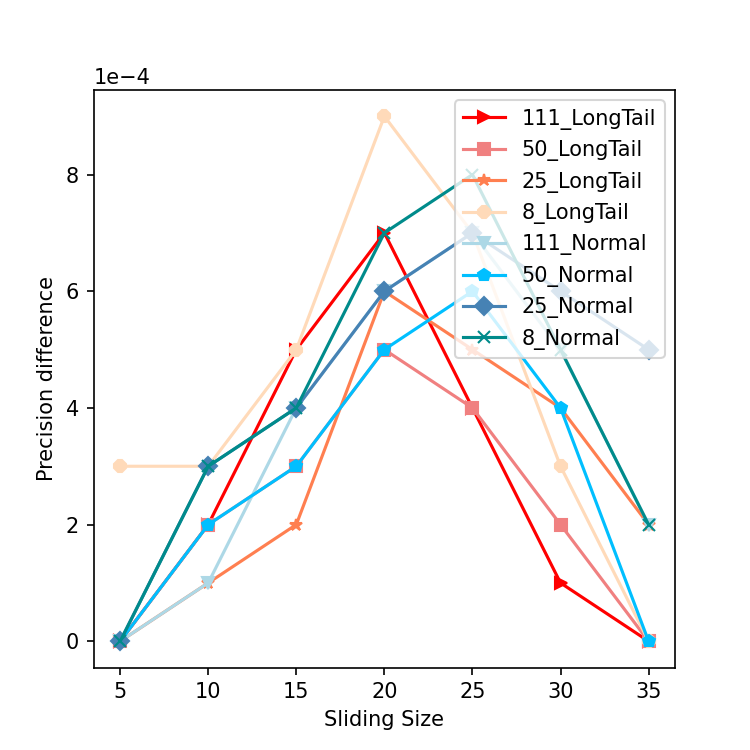}
    \label{fig:hyperparameter:sliding_size}
    }
    \subfigure[Impact of $|B_{m,i}|$.]{
    \includegraphics[width=0.23\textwidth]{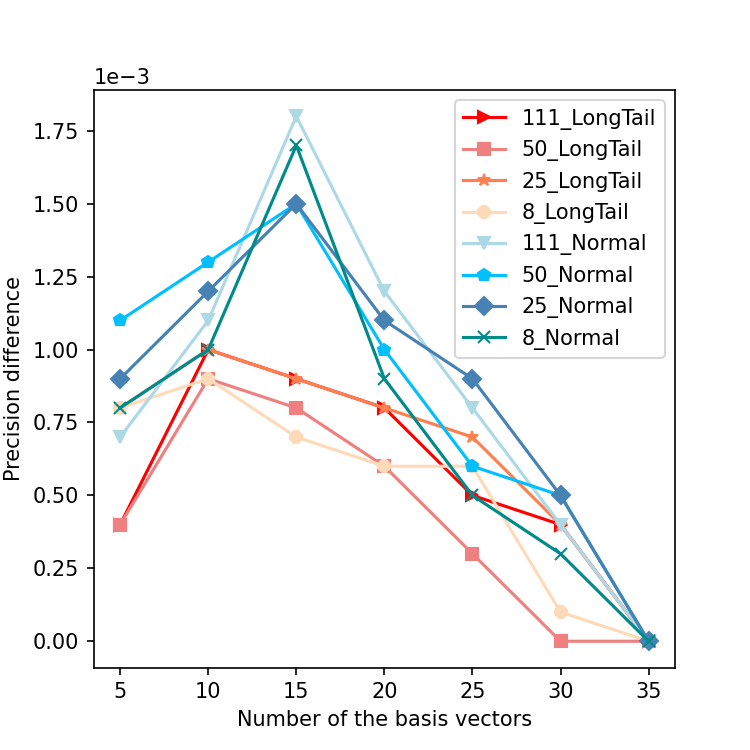}
    \label{fig:hyperparameter:number_of_b_m_i}
    }
    \subfigure[Impact of $c$.]{
    \includegraphics[width=0.23\textwidth]{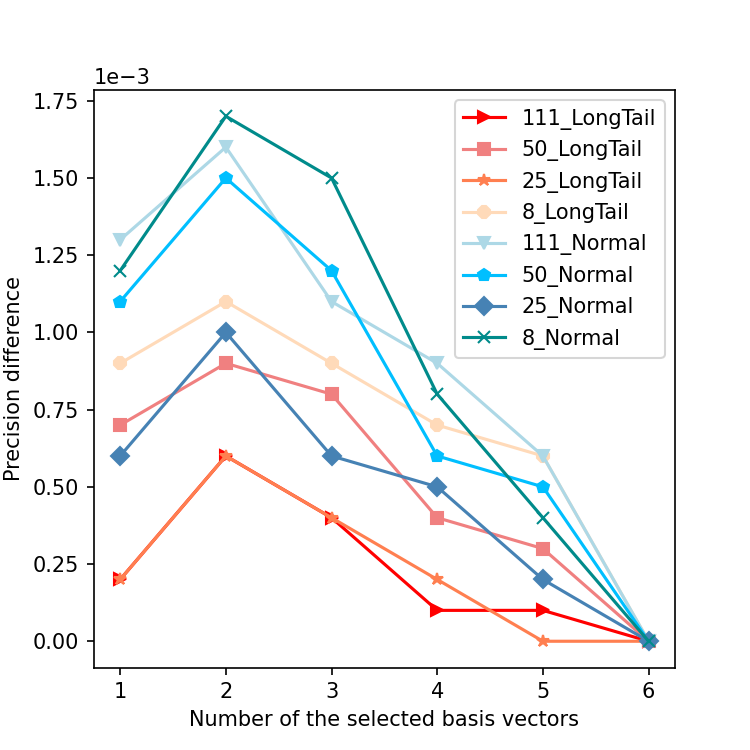}    \label{fig:hyperparameter:number_of_selected_b_m_i}
    }
    
    \subfigure[Impact of $\alpha$.]{
    \includegraphics[width=0.23\textwidth]{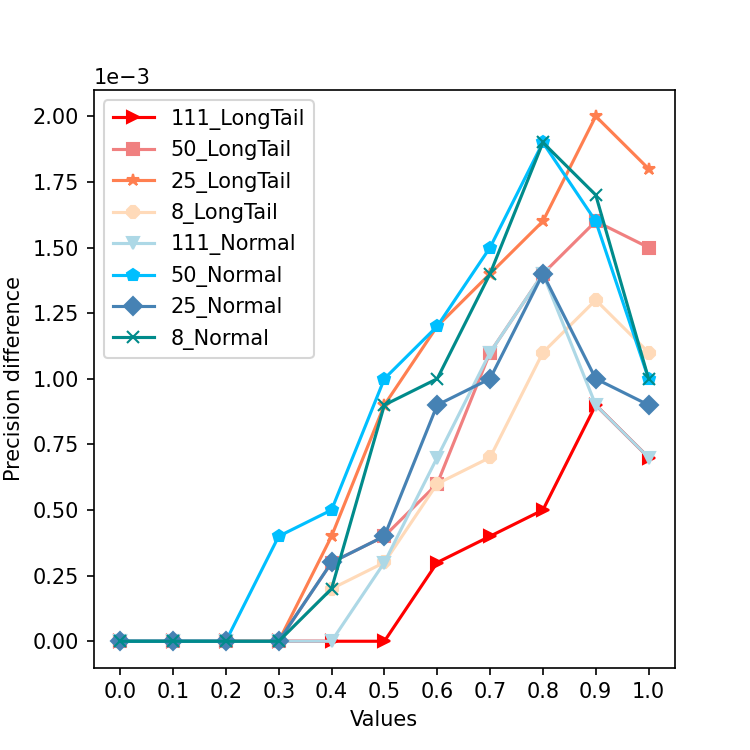}
    \label{fig:hyperparameter:alpha}
    }
    \subfigure[Impact of $\beta$.]{
    \includegraphics[width=0.23\textwidth]{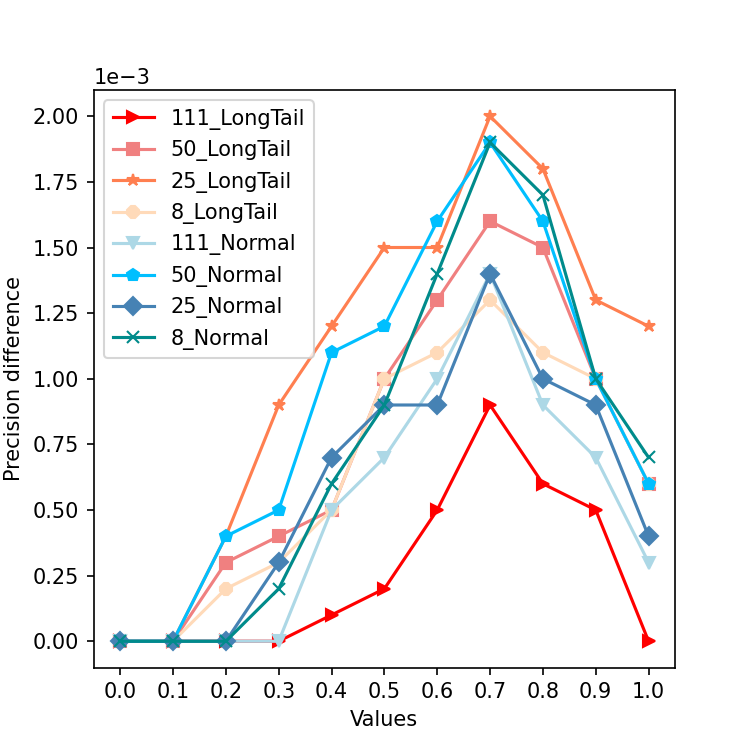}
    \label{fig:hyperparameter:belta}
    }
    \subfigure[Impact of graph depth $d$.]{
    \includegraphics[width=0.23\textwidth]{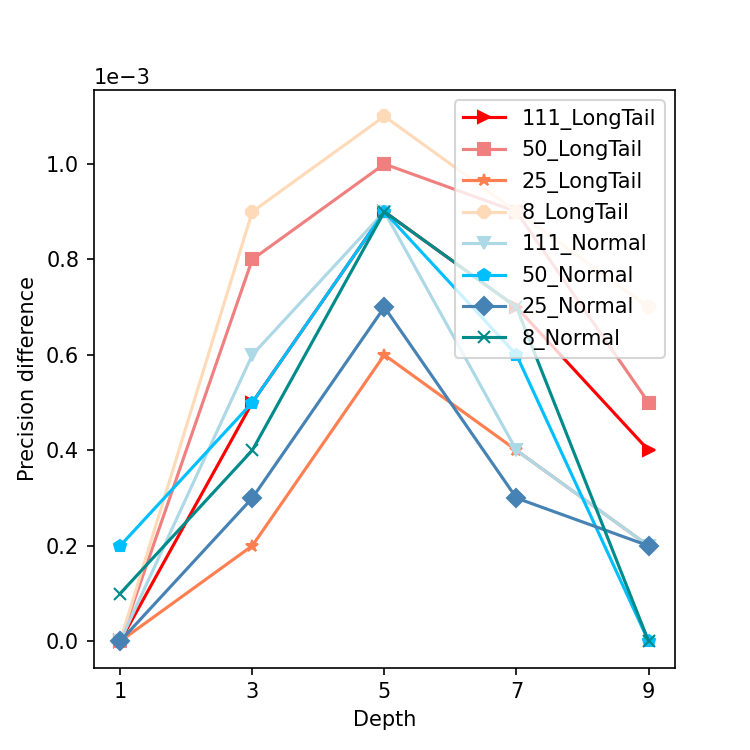}
    \label{fig:hyperparameter:graph_depth_d}
    }
    \subfigure[Impact of $\xi$.]{
    \includegraphics[width=0.23\textwidth]{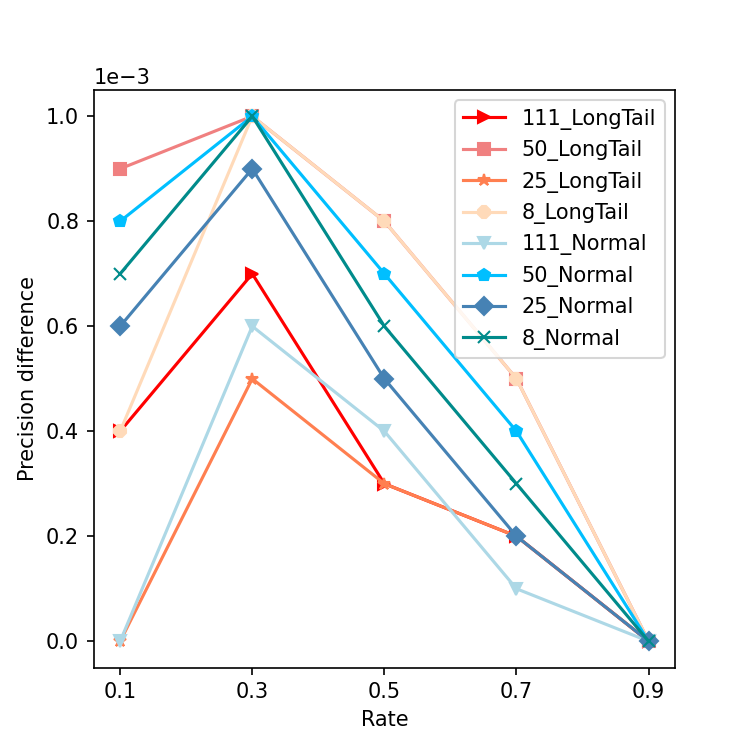}
    \label{fig:hyperparameter:xi}
    }
\caption{Plots of hyperparameter analysis. (Note: Axes are labeled in scientific notation.)}
\label{fig:hyperparameters}
\end{figure*}

The filter ratio, $\alpha$, decides how many weights should be kept in the continuity adjacency matrix. 
As it is applied to the ReLU function, some weights on the matrices are 0. 
In Figure~\ref{fig:hyperparameter:alpha}, horizontal lines are shown when the ratios are relatively small because zeros are filtered out with ratio $\alpha$.
Thus, there is no influence on model performance. 
The suitable values of filter ratios for long-tail distribution and normal distribution are 0.9 and 0.8, respectively.
When the values are smaller than the suitable values, some weights revealing irrelevant connections of two events are still used in the learning process. 
When the values are bigger than the suitable values, not enough connections are kept to learn data patterns.
Similarly, the filter ratio, $\beta$, decides how many weights should be kept in the interference adjacency matrix showing the same trends in Figure~\ref{fig:hyperparameter:belta}. 
The suitable values of filter ratios for long-tail distribution and normal distribution are 0.7, which is a trade-off between more connections and more accurate information.

\subsubsection{Impacts of graph depth $d$.}
In Figure~\ref{fig:hyperparameter:graph_depth_d}, there is one suitable value, $d=5$, for the graph convolutional network. 
From 1 to 5, the convolutional depth is not deep enough so some high-order relations and information are not well-studied. 
From 5 to 9, the deep graph convolutional network makes the representation over-smooth. 

\subsubsection{Impacts of the ratio $\xi$.}
The value of $\xi $ controls the portion of original EEG information $x^{\circ}$ when the EEG data is involved in graph convolutional computing. 
The suitable value is 0.3 for both distributions.
When the value is smaller than 0.3, more neighbor information is added to the original EEG information so the original EEG information loses its unique data pattern.
On the other hand, when the value is from 0.3 to 0.9, too much original EEG information is kept while losing the advantage of capturing the influence of neighbor events. 

\subsubsection{Conclusion of hyperparameter analysis.}
We present a detailed analysis of each tunable hyperparameter. 
In general, QUARK is not a parameter-sensitive model, because the magnitude of the precision difference is 1e-3 to 1e-4.
Even though the changes in parameters influence the performance of QUARK, it is not the main reason that our model performs better than baseline models in the comparison section~\ref{Performance_Comparison}.

\section{Metrics for Feeling and Style Detection} 
\label{Metrics:Feeling_and_Style_Detection}

\subsubsection{Content Similarity}
The learned representations of images from the image recognition model, ViTAE2, can effectively represent the image contents. So, the similarity between the representation of the original image $y_x$ (a seen image whose corresponding EEG signal is learned by QUARK) and the representation of a recommended image $y_k$ indicates how similar the image content is.
The averaged content similarity for all test cases is computed by Eq.~\eqref{eq:content_sim}.
To easily filter the recommended images whose similarity scores are lower than a threshold, the Frobenius normalization strategy is considered to scale the values.
\begin{equation}
\label{eq:content_sim}
ConSim = \frac{1}{|X|}\sum_{x \in  X}( \frac{1}{k}\sum_{i\leq k} \frac{y_{x}}{\left \| y_{x} \right \|_F} \frac{(y_{k})^T}{\left \| y_{k} \right \|_F})
\end{equation}

\subsubsection{Color Similarity}
An image is composed of three colors, red, green, and blue. Color histogram distribution similarity method~\cite{novak1992anatomy} compares the distance of two pixels from unlearned image matrices
, $y_{x}^{raw}\in \mathbb{R}^{d\times d}$ and $y_{k}^{raw}\in \mathbb{R}^{d\times d}$. 
If the distance of two pixels is large, then there is a small similarity. 
The averaged color similarity for all test cases is given in Eq.~\eqref{eq:color_sim}.
\begin{equation}
\label{eq:color_sim}
CorSim = \frac{1}{|X|}\sum_{x \in  X}\left \{ \frac{1}{k}\sum_{i\leq k} \sum_{i\leq d}  \sum_{j\leq d}(1- \frac{|(y_{x}^{raw})_{i,j}-(y_{k}^{raw})_{i,j}|}{max((y_{x}^{raw})_{i,j}, (y_{k}^{raw})_{i,j}))}) \right \}
\end{equation}

\subsubsection{Structural Similarity}
Edge detection is a simple but useful way to explore the structure information of an image. 
Following Vincent's work~\cite{vincent2009descriptive}, a white-black image that only contains detected edges with white color (value 255) and non-edges with black color (value 0) can be generated. 
For the unlearned images $y_{x}^{raw}\in \mathbb{R}^{d\times d}$ and $y_{k}^{raw}\in \mathbb{R}^{d\times d}$, edge detection method generates $y_{x}^{edge}$ and $y_{k}^{edge}$.
Then, the exclusive OR operation is applied to find the non-overlapped areas, and the summation of the non-overlapped area is normalized by the total area to range $[0,1]$.
The averaged structural similarity is obtained in Eq~\eqref{eq:structure_sim}.
\begin{equation}
\label{eq:structure_sim}
StrcSim = \frac{1}{|X|}
\sum_{x \in  X}\left \{ \frac{1}{k}
\sum_{i\leq k} 
(1-\frac{sum(XOR(y_{x}^{edge},y_{k}^{edge}))}{d\times d})
\right \}
\end{equation}

\end{sloppypar}
\end{document}